\documentclass[]{JHEP3}
\usepackage{epsfig}

\title{Oscillons in dilaton-scalar theories}

\author{Gyula Fodor\\
MTA RMKI, H-1525 Budapest 114, PO..Box 49, Hungary,

\email{gfodor@rmki.kfki.hu}}
\author{P\'eter Forg\'acs\\
MTA RMKI, H-1525 Budapest 114, P.O.Box 49, Hungary\\
LMPT, CNRS-UMR 6083, Universit\'e de Tours,\\
Parc de Grandmont, 37200 Tours, France}
\author{Zal\'an Horv\'ath and M\'ark Mezei\\
Institute for Theoretical Physics, E\"otv\"os University,\\
H-1117 Budapest, P\'azm\'any P\'eter s\'et\'any 1/A, Hungary}

\abstract{It is shown by both analytical methods and numerical
  simulations that extremely long living spherically symmetric
  oscillons appear in virtually any real scalar field theory coupled
  to a massless dilaton (DS theories).  In fact such
  ''dilatonic'' oscillons are already present in the simplest
  non-trivial DS theory -- a free massive scalar field coupled to the
  dilaton.  It is shown that in analogy to the previously considered
  cases with a single nonlinear scalar field, in DS theories there are
  also time periodic quasibreathers (QB) associated to small
  amplitude oscillons. Exploiting the QB picture the radiation law of
  the small amplitude dilatonic oscillons is determined analytically.

}


\preprint{...}

\begin{document}

%\maketitle

\section{Introduction}\label{s:intro}
Long-living, spatially localized classical solutions in field theories
containing scalar fields exhibiting nearly periodic oscillations in
time -- {\sl oscillons} -- \cite{dashen}--\cite{sicilia} have
attracted considerable interest in the last few years.  Oscillons
closely resemble ''true'' breathers of the one-dimensional ($D=1$)
sine-Gordon (SG) theory, which are time periodic and are exponentially
localized in space, but unlike true breathers they are continuously
losing energy by radiating slowly.  On the other hand oscillons exist
for different scalar potentials in various spatial dimensions, in
particular for $D=1,2,3$.  Just like a breather, an oscillon possesses
a spatially well localized ``core'', but it also has a ``radiative''
region outside of the core. Oscillons appear from rather generic
initial data in the course of time evolution in an impressive number
of physically relevant theories including the bosonic sector of the
standard model \cite{Farhi05}--\cite{Borsanyi2}. Moreover they form in
physical processes making them of considerable importance
\cite{Kolb:1993hw}--\cite{GleiserTHor08}.  In a series of papers,
\cite{FFGR}, \cite{FFHL}--\cite{moredim}, it has been shown that
oscillons can be well described by a special class of exactly
time-periodic ''quasibreathers'' (QB). QBs also possess a well
localized core in space (just like true breathers) but in addition
they have a standing wave tail whose amplitude is minimized.  At this
point it is important to emphasize that there are (infinitely) many
time periodic solutions characterized by an asymptotically standing
wave part.  In order to select one solution, we impose the condition
that the standing wave amplitude be minimal.  This is a physically
motivated condition, which heuristically should single out ''the''
solution approximating a true breather as well as possible, for which
this amplitude is identically zero.  The amplitude of the standing
wave tail of a QB is closely related to that of the oscillon
radiation, therefore its computation is of prime interest.  It is a
rather non-trivial problem to compute this amplitude even in one
spatial dimensional scalar theories \cite{SK}, \cite{FFHM}.  In the
limit when the core amplitude is small, we have developed a method to
compute the leading part of the exponentially suppressed tail
amplitude for a general class of theories in various dimensions
\cite{moredim}.

In this paper we show that oscillons also appear in rather general
(real) scalar field theories coupled to a (massless) dilaton field (DS
theory).  Dilaton fields appear naturally in low energy effective
field theories derived from superstring models \cite{Wi, BFQ, GSW} and
the study of their effects is of major interest.  As the present study
shows, the coupling of a dilaton even to a free massive scalar field,
referred to as the dilaton-Klein-Gordon (DKG) theory, which is the
conceivably simplest non-trivial DS theory, has some rather remarkable
consequences.  This simple DKG theory already admits QBs and as our
numerical investigations show from generic initial data small
amplitude oscillons evolve.  We concentrate on solutions with the
simplest spatial geometry - spherical symmetry. We do not think that
considering spherically symmetric configurations is a major
restriction since non-symmetric configurations are expected to contain
more energy and to evolve into symmetric ones
\cite{Hindmarsh-Salmi07}.  The dilatonic oscillons are very robust and
once formed from the initial data they do not even seem to radiate
their energy, hence their lifetime is extremely long (not even
detectable by our numerical methods).

Our means for constructing dilatonic oscillons will be the small
amplitude expansion, in which the small parameter, $\varepsilon$,
determines the difference of oscillation frequency from the mass
threshold. The small amplitude oscillons of the DKG theory appear to
be stable in dimensions $D=3,4$, unstable in $D=5,6$, and their core
amplitude is proportional to $\varepsilon^2$.  This is to be
contrasted to self-interacting scalar theories whose oscillons are
stable in $D=1,2$, unstable in $D=3$, and their core amplitude is
proportional to $\varepsilon$. The master equations determining
oscillons to leading order in the small amplitude expansion turn out
to be the Schr\"odinger-Newton (SN) equations.  The main analytical
result of this paper is the analytic computation of the amplitude of
the standing wave tail of the dilatonic QBs for any dimension $D$, and
thereby the determination of the radiation law and the lifetime of
small amplitude oscillons in DS theories. The used methods have been
developed in Refs.\ \cite{SK}, \cite{Kichenassamy}, \cite{Pomeau},
\cite{FFHM} and \cite{moredim}.

The above results, namely the stability properties and the SN
equations playing the r\^ole of master equation, show striking
similarity to those obtained in the Einstein-Klein-Gordon (EKG)
theory, i.e.\ for a free massive scalar field coupled to Einstein's
gravity, where also stable, long living oscillons (known under the
name of oscillating soliton stars, or more recently oscillatons) have
been found and investigated in many papers
\cite{Seidel}--\cite{Kichena2}.

\section{The scalar-dilaton system}

The action of a scalar-dilaton system is
\begin{equation}\label{action}
A=\int dt\, d^D\! x \left[\frac12(\partial_\mu\varphi)^2
+\frac12(\partial_\mu\Phi)^2
-e^{-2\kappa \varphi}\,U(\Phi)\right] \,,
\end{equation}
where $\varphi$ is the dilaton field and $\Phi$ is a scalar field with
self interaction potential $U(\Phi)$.

The energy corresponding to the action (\ref{action}) can be written
as
\begin{equation}\label{e:density}
E = \int d^{D}x\,{\cal E}\,, \qquad
{\cal{E}} = \frac{1}{2}\left[
\left(\partial_t \Phi\right)^2
+\left(\partial_i \Phi\right)^2
+\left(\partial_t \varphi\right)^2
+\left(\partial_i \varphi\right)^2
\right]
+e^{-2\kappa \varphi}U(\Phi)\,,
\end{equation}
where ${\cal{E}}$ denotes the energy density. In the case of spherical
symmetry
\begin{equation}\label{e:energysph}
E = \int_0^\infty dr\,\frac{\pi^{D/2}r^{D-1}}{\Gamma(D/2)}\left[
\left(\partial_t \Phi\right)^2
+\left(\partial_r \Phi\right)^2
+\left(\partial_t \varphi\right)^2
+\left(\partial_r \varphi\right)^2
+2 e^{-2\kappa \varphi}U(\Phi)
\right] \,.
\end{equation}

We assume that the potential
can be expanded around its minimum at $\Phi=0$ as
\begin{equation}
U(\Phi)=\sum_{k=1}^{\infty}\frac{g_k}{k+1}\Phi^{k+1} \,, \qquad
U'(\Phi)=\sum_{k=1}^{\infty}g_k\Phi^{k} \ ,
\end{equation}
where $g_k$ are real constants. For a free massive scalar field with
mass $m$ the only nonzero coefficient is $g_1=m^2$. If $g_{2k}=0$ for
integer $k$ the potential is symmetric around its minimum. In that
case, as we will see, for periodic configurations the Fourier
expansion of $\Phi$ in $t$ will contain only odd, while the expansion
of $\varphi$ only even Fourier components. For spherically symmetric
systems the field equations are
\begin{eqnarray}
-\frac{\partial^2\Phi}{\partial t^2}
+ \frac{\partial^2\Phi}{\partial r^2}
+\frac{D-1}{r}\,\frac{\partial\Phi}{\partial r} &=&
e^{-2\kappa \varphi}\,U'(\Phi) \,,
\label{evolPhi}\\
-\frac{\partial^2\varphi}{\partial t^2}
+\frac{\partial^2\varphi}{\partial r^2}
+\frac{D-1}{r}\,\frac{\partial\varphi}{\partial r} &=&
-2\kappa\,e^{-2\kappa \varphi}\, U(\Phi)\,.\label{evolvarphi}
\end{eqnarray}
Since $g_1=m$ is intended to be the mass of small excitations of
$\Phi$ at large distances, we look for solutions satisfying
$\varphi\to 0$ for $r\to\infty$. Finiteness of energy also requires
$\Phi\to 0$ as $r\to\infty$. Rescaling the coordinates as $t\to t/m$
and $r\to r/m$ we first set $g_1=m^2=1$. Then redefining
$\varphi\to\varphi/(2\kappa)$ and $\Phi\to\Phi/(2\kappa)$ and
appropriately changing the constants $g_k$ we arrange that
$2\kappa=1$. If for some reason we obtain a solution for which
$\varphi$ tends to a nonzero constant at infinity then the dilatation
symmetry of the system allows us to shift $\varphi$ and rescale the
coordinates so that it is transformed to a solution satisfying
$\varphi\to 0$ for $r\to\infty$.

An important feature of a localized dilatonic configuration is its
dilaton charge, $Q$. It can be defined for almost time-periodic
spherically symmetric configurations like oscillons as:
\begin{eqnarray}
	\varphi&\approx& Q\,r^{2-D}  \qquad {\rm for}\;
r\to\infty \;{\rm in}\; D\not=2 \label{eq:charge}\\
	\varphi&\approx& Q\,\ln r \qquad{\rm for}\;
r\to\infty \;{\rm in}\; D=2\,.
\end{eqnarray}

\section{The small amplitude expansion}\label{sec:smallampl}

In this section we will construct a finite-energy family of localized
small amplitude solutions of the spherically symmetric field equations
(\ref{evolPhi}) and (\ref{evolvarphi}) which oscillate below the mass
threshold \cite{Kichenassamy}. It will be shown that such solutions
exist for $2<D<6$. The subtleties of the case $D=6$ will be dealt with
in subsection \ref{sec:dsix}. The result of the small amplitude
expansion is an asymptotic series representation of the core region of
a quasibreather or oscillon, but misses a standing or outgoing wave
tail whose amplitude is exponentially small with respect to the
core. The amplitude of the tail will be determined in section
\ref{s:borel}.

We are looking for small amplitude solutions, therefore we expand the
scalar fields, $\varphi$ and $\Phi$, in terms of a parameter
$\varepsilon$ as
\begin{equation}
\varphi=\sum_{k=1}^\infty\varepsilon^k \varphi_k \,,\qquad
\Phi=\sum_{k=1}^\infty\varepsilon^k \Phi_k
\,, \label{sumphi}
\end{equation}
and search for functions $\phi_k$ and $\Phi_k$ tending to zero at
$r\to\infty$.  The size of smooth configurations is expected to
increase for decreasing values of $\varepsilon$, therefore it is
natural to introduce a new radial coordinate by the following
rescaling
\begin{equation}
\rho=\varepsilon r \,.
\end{equation}
In order to allow for the $\varepsilon$ dependence of the time-scale
of the configurations a new time coordinate is introduced as
\begin{equation}
\tau=\omega(\varepsilon) t \,.
\end{equation}
Numerical experience shows that the smaller the oscillon amplitude is
the closer its frequency becomes to the threshold $\omega=1$.  The
function $\omega(\varepsilon)$ is assumed to be analytic near
$\omega=1$, and it is expanded as
\begin{equation}
\omega^2(\varepsilon)=1+\sum_{k=1}^\infty\varepsilon^k\omega_k \,.
\end{equation}
We note that there is a considerable freedom in choosing different
parametrisations of the small amplitude states, changing the actual
form of the function $\omega(\varepsilon)$. The physical parameter is
not $\varepsilon$ but the frequency of the periodic states that will be
given by $\omega$.  After the rescalings Eqs.\ (\ref{evolPhi}) and
(\ref{evolvarphi}) take the following form
\begin{eqnarray}
-\omega^2\frac{\partial^2\Phi}{\partial \tau^2}
+\varepsilon^2\frac{\partial^2\Phi}{\partial\rho^2}
+\varepsilon^2\frac{D-1}{\rho}\,\frac{\partial\Phi}{\partial\rho}
&=& e^{-\varphi}\left(\Phi+\sum_{k=2}^{\infty}g_k\Phi^{k}
\right) ,\label{evolPhi2}\\
-\omega^2\frac{\partial^2\varphi}{\partial \tau^2}
+\varepsilon^2\frac{\partial^2\varphi}{\partial\rho^2}
+\varepsilon^2\frac{D-1}{\rho}\,\frac{\partial\varphi}{\partial\rho}
&=& -e^{-\varphi}\left(\frac{1}{2}\Phi^2
+\sum_{k=2}^{\infty}\frac{g_k}{k+1}\Phi^{k+1}\right)
.\label{evolvarphi2}
\end{eqnarray}

Substituting the small amplitude expansion (\ref{sumphi}) into
(\ref{evolPhi2}) and (\ref{evolvarphi2}), to leading $\varepsilon$
order we obtain
\begin{equation}
\frac{\partial^2\Phi_1}{\partial \tau^2}+\Phi_1=0 \,, \qquad
\frac{\partial^2\varphi_1}{\partial \tau^2}=0 \,. \label{eps1ord}
\end{equation}
Since we are looking for solutions which remain bounded in time and
since we are free to shift the origin $\tau=0$ of the time coordinate,
the solution of (\ref{eps1ord}) can be written as
\begin{equation}
\Phi_1(\tau,\rho)=P_1(\rho)\cos \tau \,, \qquad
\varphi_1(\tau,\rho)=p_1(\rho) \,,
\end{equation}
where $P_1(\rho)$ and $p_1(\rho)$ are some functions of the rescaled
radial coordinate $\rho$.

The $\varepsilon^2$ terms in the expansion of (\ref{evolvarphi2})
yield
\begin{equation}
\frac{\partial^2\varphi_2}{\partial \tau^2}=
\frac{1}{4}P_1^2\left[1+\cos(2\tau)\right] \,. \label{eps2varphi}
\end{equation}
This equation can have a solution for $\varphi_2$ which remains
bounded in time only if the time independent term in the right hand
side vanishes, implying $P_1=0$ and consequently $\Phi_1=0$. Then the
solution of (\ref{eps2varphi}) is
$\varphi_2(\tau,\rho)=p_2(\rho)$. The $\varepsilon^2$ terms in
(\ref{evolPhi2}) yield
\begin{equation}
\frac{\partial^2\Phi_2}{\partial \tau^2}+\Phi_2=0 \,.
\end{equation}
Since $\Phi_1=0$ we are again free to shift the time coordinate, and
the solution is $\Phi_2(\tau,\rho)=P_2(\rho)\cos \tau$.

The $\varepsilon^3$ order terms in the expansion of
(\ref{evolvarphi2}) give
\begin{equation}
\frac{\partial^2\varphi_3}{\partial \tau^2}=
\frac{d^2p_1}{d\rho^2}
+\frac{D-1}{\rho}\,\frac{d p_1}{d\rho}
 \,. \label{eps3varphi}
\end{equation}
In order to have a solution for $\varphi_3(\tau,\rho)$ that remains
bounded in time, the right hand side must be zero, yielding
$p_1(\rho)=p_{11}+p_{12}\rho^{2-D}$ when $D\not=2$ and
$p_1(\rho)=p_{11}+p_{12}\ln\rho$ for $D=2$, with some constants
$p_{11}$ and $p_{12}$. Since we look for bounded regular solutions
tending to zero at $\rho\to\infty$, we must have $p_{11}=p_{12}=0$. As
we have already seen that $\Phi_1=0$, this means that the small
amplitude expansion (\ref{sumphi}) starts with $\varepsilon^2$
terms. The solution of (\ref{eps3varphi}) is then
$\varphi_3(\tau,\rho)=p_3(\rho)$.  The $\varepsilon^3$ order terms in
the expansion of (\ref{evolPhi2}) give
\begin{equation}
\frac{\partial^2\Phi_3}{\partial \tau^2}+\Phi_3
-\omega_1 P_2\cos\tau=0 \,. \label{eqPhi3}
\end{equation}
This equation can have a solution for $\Phi_3$ which remains bounded
in time only if the resonance term proportional to $\cos \tau$ vanishes,
implying $\omega_1=0$.  After applying an $\varepsilon^3$ order small
shift in the time coordinate, the solution of (\ref{eqPhi3}) is
$\Phi_3(\tau,\rho)=P_3(\rho)\cos \tau$. Continuing to higher orders,
the basic frequency $\sin \tau$ term can always be absorbed by a small
shift in $\tau$. It is important to note that after transforming out
the $\sin \tau$ terms no $\sin(k\tau)$ terms will appear in the
expansion, implying the time reflection symmetry of $\Phi$ and
$\varphi$ at $\tau=0$.

\subsection{The Schr\"odinger-Newton equations}\label{sec:sn}

The $\varepsilon^4$ terms in the expansion of (\ref{evolPhi2}) and
(\ref{evolvarphi2}) yield the differential equations
\begin{eqnarray}
\frac{\partial^2\Phi_4}{\partial \tau^2}+\Phi_4&=&
\left[
\frac{d^2P_2}{d\rho^2}
+\frac{D-1}{\rho}\,\frac{d P_2}{d\rho}
+(p_2+\omega_2) P_2
\right]\cos \tau
-\frac{1}{2}g_2P_2^2\left[1+\cos(2\tau)\right]
, \label{eps4Phi}\\
\frac{\partial^2\varphi_4}{\partial \tau^2}&=&
\frac{d^2p_2}{d\rho^2}
+\frac{D-1}{\rho}\,\frac{d p_2}{d\rho}
+\frac{1}{4}P_2^2\left[1+\cos(2\tau)\right]
. \label{eps4varphi}
\end{eqnarray}
The function $\Phi_4(\tau,\rho)$ and $\varphi_4(\tau,\rho)$ can remain
bounded only if the $\cos \tau$ resonance terms in (\ref{eps4Phi}) and
the time independent terms in (\ref{eps4varphi}) vanish,
\begin{eqnarray}
\frac{d^2P_2}{d\rho^2}
+\frac{D-1}{\rho}\,\frac{d P_2}{d\rho}
+(p_2+\omega_2) P_2&=&0
\,, \label{P2eq}\\
\frac{d^2p_2}{d\rho^2}
+\frac{D-1}{\rho}\,\frac{d p_2}{d\rho}
+\frac{1}{4}P_2^2&=&0
\,. \label{p2eq}
\end{eqnarray}
Then the time dependence of $\Phi_4(\tau,\rho)$ and
$\varphi_4(\tau,\rho)$ is determined by (\ref{eps4Phi}) and
(\ref{eps4varphi}) as
\begin{equation}
\Phi_4(\tau,\rho)=P_4(\rho)\cos \tau
+\frac{1}{6}g_2P_2^2\left[\cos(2\tau)-3\right]\,,\qquad
\varphi_4(\tau,\rho)=p_4(\rho)-\frac{1}{16}P_2(\rho)^2\cos(2\tau)\,.
\end{equation}
Here we see the first contribution of a nontrivial $U(\Phi)$
potential, the term proportional to $g_2$ in $\Phi_4$. If (and only
if) the potential is non-symmetric around its minimum, even Fourier
components appear in the expansion of $\Phi$.

Introducing the new variables
\begin{equation}
S=\frac{1}{2}P_2 \,, \qquad s=p_2+\omega_2 \,, \label{eq:defss}
\end{equation}
(\ref{P2eq}) and (\ref{p2eq}) can be written into the form which is
called the time-independent Schr\"odinger-Newton (or
Newton-Schr\"odinger) equations in the literature:
\begin{eqnarray}
\frac{d^2S}{d\rho^2}
+\frac{D-1}{\rho}\,\frac{d S}{d\rho}
+s S&=&0
\,, \label{Seq}\\
\frac{d^2s}{d\rho^2}
+\frac{D-1}{\rho}\,\frac{d s}{d\rho}
+S^2&=&0
\,. \label{seq}
\end{eqnarray}
We look for localized solutions of these equations, in order to
determine the core part of small amplitude oscillons to a leading
order approximation in $\varepsilon$.  The main features of the
solutions depend on the number of spatial dimensions $D$.  For $D\geq
6$ positive monotonically decreasing solutions necessarily satisfy
$s=S$, they tend to zero, furthermore, the Lane-Emden
equation holds \cite{choquard}
\begin{equation}
\frac{d^2s}{d\rho^2}
+\frac{D-1}{\rho}\,\frac{d s}{d\rho}+s^2=0 \,.\label{emden}
\end{equation}
For $D>6$ solutions are decreasing as $1/\rho^2$ for large $\rho$,
consequently they have infinite energy. It can also be shown that
solutions of the original Schr\"odinger-Newton system with $s\not=S$,
and a necessarily oscillating scalar field, have infinite energy,
hence there is no finite energy solution for $D>6$. For $D=6$ the
explicit form of the asymptotically decaying solutions of
(\ref{emden}) are known
\begin{equation}
s=\pm S=\frac{24\alpha^2}{\left(1+\alpha^2\rho^2\right)^2} \,,
\label{6dss}
\end{equation}
where $\alpha$ is any constant. Since the replacement of $\Phi$ with
$-\Phi$ and a simultaneous reflection of the potential around its
minimum is a symmetry of the system, we choose the positive sign for
$S$ in (\ref{6dss}).  For $D=6$ the total energy remains finite.

If $D<6$, then localized solutions have the property that for large
values of $\rho$ the function $S$ tends to zero exponentially, while
$s$ behaves as $s\approx s_0+s_1\rho^{2-D}$ for $D\not=2$ and as
$s\approx s_0+s_1\ln\rho$ for $D=2$, where $s_0$ and $s_1$ are some
constants. Since we are interested in localized solutions we assume
$2<D<6$.  From (\ref{Seq}) it is apparent that exponentially localized
solutions for $S$ can only exist if $s$ tends to a negative constant,
i.e. $s_0<0$.  In this case the localized solutions of the
Schr\"odinger-Newton (SN) equations (\ref{Seq}) and (\ref{seq}) can be
parametrized by the number of nodes of $S$. The physically important
ones are the nodeless solutions satisfying $S>0$, since the others
correspond to higher energy and less stable oscillons.

Motivated by the asymptotic behaviour of $s$, if $D\not=2$ it is
useful to introduce the variables
\begin{equation}
\mu=\frac{\rho^{D-1}}{2-D}\,\frac{d s}{d\rho} \,, \qquad
\nu=s-\rho^{2-D}\mu \ .
\end{equation}
In $2<D<6$ dimensions these variables tend exponentially to the
earlier introduced constants
\begin{equation}
\lim_{\rho\to\infty}\mu=s_1 \,, \qquad
\lim_{\rho\to\infty}\nu=s_0 \,. \label{e:s0s1}
\end{equation}
Then the SN equations can be written into the equivalent form
\begin{eqnarray}
\frac{d\mu}{d\rho}+\frac{\rho^{D-1}}{2-D}S^2&=&0 \,,\label{e:dmu}\\
\frac{d\nu}{d\rho}+\frac{\rho}{D-2}S^2&=&0 \,,\\
\frac{d^2S}{d\rho^2}
+\frac{D-1}{\rho}\,\frac{d S}{d\rho}
+\left(\nu+\rho^{2-D}\mu\right) S&=&0 \,.
\end{eqnarray}

The SN equations (\ref{Seq}) and (\ref{seq}) have the scaling
invariance
\begin{equation}
(S(\rho),s(\rho)) \to
(\lambda^2S(\lambda\rho),\lambda^2s(\lambda\rho))
\,.\label{snscale}
\end{equation}
If $2<D<6$ we use this freedom to make the nodeless solution unique by
setting $s_0=-1$. At the same time we change the $\varepsilon$
parametrization by requiring
\begin{equation}
\omega_2=-1 \ \  {\rm for} \ \  2<D<6 \,,
\end{equation}
ensuring that the limiting value of $\varphi$
vanishes to $\varepsilon^2$ order. Going to higher orders, it can be
shown that one can always make the choice $\omega_i=0$ for $i\geq 3$,
thereby fixing the $\varepsilon$ parametrization, and setting
\begin{equation}\label{eq:omeps}
\omega=\sqrt{1-\varepsilon^2} \ \  {\rm for} \ \  2<D<6 \,.
\end{equation}
If $D=6$, since both $s$ and $S$ tend to zero at infinity, we have no
method yet to fix the value of $\alpha$ in (\ref{6dss}). Moreover, in
order to ensure that $\varphi$ tends to zero at infinity we have to
set
\begin{equation}
\omega_2=0 \ \  {\rm for} \ \  D=6 \,.
\end{equation}

\subsection{Absence of odd $\varepsilon$ powers in the expansion}

Calculating the $\varepsilon^5$ order equations from (\ref{evolPhi2})
and (\ref{evolvarphi2}) and requiring the boundedness of $\Phi_5$ and
$\varphi_5$ we obtain a pair of equations for $P_3$ and $p_3$:
\begin{eqnarray}
\frac{d^2P_3}{d\rho^2}
+\frac{D-1}{\rho}\,\frac{d P_3}{d\rho}
+(p_2+\omega_2) P_3+P_2 (p_3+\omega_3) &=&0 \,,\label{P3eq}\\
\frac{d^2p_3}{d\rho^2}
+\frac{D-1}{\rho}\,\frac{d p_3}{d\rho}
+\frac{1}{2}P_2P_3&=&0 \,.\label{p3eq}
\end{eqnarray}
These equations are solved by constant multiples of
\begin{equation}
P_3=2P_2+\rho\frac{dP_2}{d\rho} \,, \qquad
p_3+\omega_3=2(p_2+\omega_2)+\rho\frac{dp_2}{d\rho} \,, \label{e:p3}
\end{equation}
corresponding to the scaling invariance (\ref{snscale}) of the SN
equations. In $D>2$ dimensions $p_3$ given by (\ref{e:p3}) tends to
$\omega_3-2$ for large $\rho$. Since we are looking for solutions for
which $\varphi$ tends to zero asymptotically, after choosing
$\omega_3=0$ we can only use the trivial solution $P_3=p_3=0$. The
important consequence is that $\Phi_3=\varphi_3=0$. Going to higher
orders in the $\varepsilon$ expansion, at odd orders we get the same
form of equations as (\ref{P3eq}) and (\ref{p3eq}), consequently, all
odd coefficients of $\Phi_k$ and $\varphi_k$ can be made to
vanish. Instead of the more general form (\ref{sumphi}) we can write
the small amplitude expansion as
\begin{equation}
\varphi=\sum_{k=1}^\infty\varepsilon^{2k} \varphi_{2k} \,,\qquad
\Phi=\sum_{k=1}^\infty\varepsilon^{2k} \Phi_{2k}
\,. \label{sumphi2}
\end{equation}

\subsection{Higher orders in the $\varepsilon$ expansion}

The $\varepsilon^6$ order equations, after requiring the boundedness
of $\Phi_6$ and $\varphi_6$, yield a pair of equations for $P_4$ and
$p_4$:
\begin{eqnarray}
&&\frac{d^2P_4}{d\rho^2}
+\frac{D-1}{\rho}\,\frac{d P_4}{d\rho}
+(p_2+\omega_2) P_4+P_2 (p_4+\omega_4) \nonumber\\
&&\qquad-\frac{1}{2}p_2^2P_2-\frac{1}{32}P_2^3
+\left(\frac{5}{6}g_2^2-\frac{3}{4}g_3\right)P_2^3
=0 \,,\label{P4eq}\\
&&\frac{d^2p_4}{d\rho^2}
+\frac{D-1}{\rho}\,\frac{d p_4}{d\rho}
+\frac{1}{2}P_2P_4-\frac{1}{4}p_2P_2^2
=0 \,.\label{p4eq}
\end{eqnarray}
This is an inhomogeneous linear system of differential equations with
nonlinear, asymptotically decaying source terms given by the solutions
of the SN equations.  Since the homogeneous terms have the same
structure as in (\ref{P3eq}) and (\ref{p3eq}), one can always add
multiples of
\begin{equation}
P_4^{(h)}=2P_2+\rho\frac{dP_2}{d\rho} \,, \qquad
p_4^{(h)}+\omega_4=2(p_2+\omega_2)+\rho\frac{dp_2}{d\rho} 
\,, \label{homp4}
\end{equation}
to a particular solution of (\ref{P4eq}) and (\ref{p4eq}).  If $2<D<6$
we are interested in solutions for which at large radii $P_4$ decays
exponentially, and $p_4\approx q_0+q_1\rho^{2-D}$ with some constants
$q_0$ and $q_1$.  We use the homogeneous solution (\ref{homp4}) to
make $q_0=0$. Since similar choice can be made at higher $\varepsilon$
orders, this will ensure that the limit of $\varphi$ will remain zero
at $\rho\to\infty$. We note that, in general, it is not possible to
make $q_1$ also vanish, implying a nontrivial $\varepsilon$ dependence
of the dilaton charge $Q$.

The resulting expressions for the original $\Phi$ and $\varphi$
functions are
\begin{eqnarray}
\Phi&=&\varepsilon^2P_2\cos\tau
+\varepsilon^4\left\{P_4\cos\tau
+\frac{1}{6}g_2P_2^2\left[\cos(2\tau)-3\right]\right\}
+\varepsilon^6\Biggl\{P_6\cos \tau
\nonumber\\
&&+\frac{P_2^3}{256}\left(1+\frac{16}{3}g_2^2+8g_3\right)\cos(3\tau)
-g_2\left[P_2P_4-(p_2+\omega_2)P_2^2
+\left(\frac{d P_2}{d\rho}\right)^2\right]\label{Phires}\\
&&+\frac{g_2}{9}\left[3P_2P_4-(p_2+\omega_2)P_2^2
-\left(\frac{d P_2}{d\rho}\right)^2\right]\cos(2\tau)
\Biggr\}+{\cal O}(\varepsilon^8) \,,\nonumber\\
\varphi&=&\varepsilon^2p_2
+\varepsilon^4\left[p_4-\frac{P_2^2}{16}\cos(2\tau)\right]
+\varepsilon^6\Biggl\{p_6-
\frac{1}{32}\negthinspace
\left[4P_2P_4-(p_2+\omega_2)P_2^2-
\left(\frac{d P_2}{d\rho}\right)^2\right]
\cos(2\tau)\nonumber\\
&&
+\frac{1}{54}g_2P_2^3\left[9\cos\tau-\cos(3\tau)\right]
\Biggr\}+{\cal O}(\varepsilon^8) \,, \label{varphires}
\end{eqnarray}
where the functions $P_2$ and $p_2$ are determined by the SN equations
(\ref{P2eq}) and (\ref{p2eq}), $P_4$ and $p_4$ can be obtained from
(\ref{P4eq}) and (\ref{p4eq}), furthermore, the equations for $P_6$
and $p_6$ can be calculated from the $\varepsilon^8$ order terms as
\begin{eqnarray}
&&\frac{d^2P_6}{d\rho^2}
+\frac{D-1}{\rho}\,\frac{d P_6}{d\rho}
+(p_2+\omega_2)P_6+(p_6+\omega_6)P_2
+\left(p_4+\omega_4-\frac{p_2^2}{2}\right)P_4\nonumber\\
&&\qquad-\left(\frac{3}{32}-\frac{5}{2}g_2^2+\frac{9}{4}g_3\right)P_2^2P_4
-p_2 P_2 p_4
+\left(\frac{3}{64}-\frac{49}{18}g_2^2+\frac{3}{4}g_3\right)p_2P_2^3
\label{P6eq}\\
&&\qquad+\left(\frac{1}{64}-\frac{17}{9}g_2^2\right)\omega_2P_2^3
+\frac{1}{6}p_2^3 P_2
+P_2\left(\frac{1}{64}+\frac{19}{9}g_2^2\right)
\left(\frac{d p_2}{d\rho}\right)^2
=0 \,,\nonumber\\
&&\frac{d^2p_6}{d\rho^2}
+\frac{D-1}{\rho}\,\frac{d p_6}{d\rho}
+\frac{1}{2}P_2P_6+\frac{1}{4}P_4^2-\frac{1}{2}p_2 P_2 P_4\nonumber\\
&&\qquad-\frac{1}{4}P_2^2 p_4+\frac{1}{8}p_2^2 P_2^2
+\frac{P_2^4}{16}\left(\frac{1}{8}-\frac{11}{9}g_2^2+\frac{3}{2}g_3
\right)=0 \,.\label{p6eq}
\end{eqnarray}
We remind the reader that the only non-vanishing $\omega_k$ for $2<D<6$
is $\omega_2=-1$, and we will show in Subsection \ref{sec:dsix}, that
in general, for $D=6$ the only nonzero component is $\omega_4$. The
above expressions, especially those for $\Phi$ and $\varphi$, simplify
considerably for symmetric $U(\Phi)$ potentials, in which case
$g_2=0$.

\subsection{Free scalar field in $2<D<6$ dimensions}

If $\Phi$ is a free massive field with potential
$U(\Phi)=m^2\Phi^2/2$, after scaling out $m$ and $\kappa$ no
parameters remain in the equations determining $P_i$ and $p_i$. The
spatially localized nodeless positive solution of the ordinary
differential equations (\ref{P2eq}), (\ref{p2eq}), and the
corresponding solution of (\ref{P4eq}), (\ref{p4eq}), (\ref{P6eq}) and
(\ref{p6eq}) can be calculated numerically.  For $D=3$ the obtained
curves are shown on Figs.~\ref{f:3dP} and \ref{f:3dp}.
\EPSFIGURE{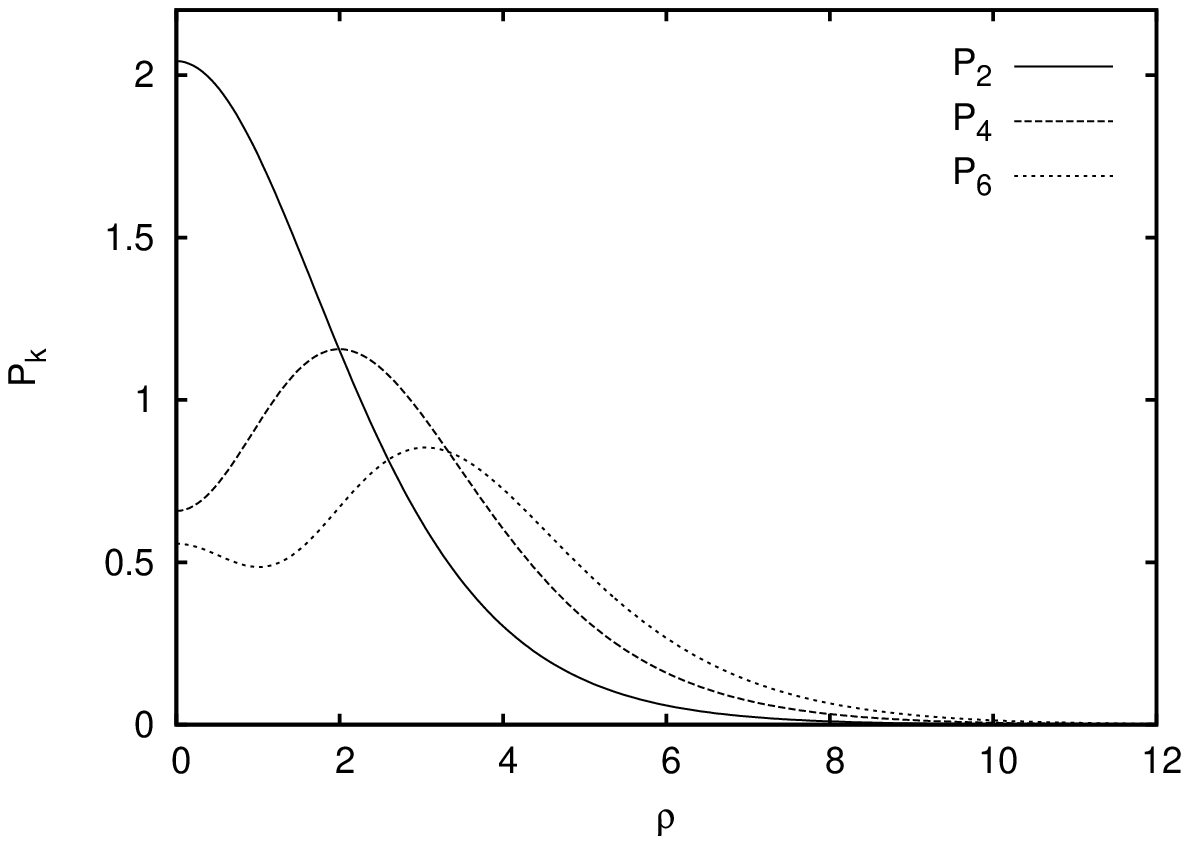,width=9.0cm} {The first three $P_k$ functions
  for the free scalar field case in $D=3$ spatial
  dimensions. \label{f:3dP}} \EPSFIGURE{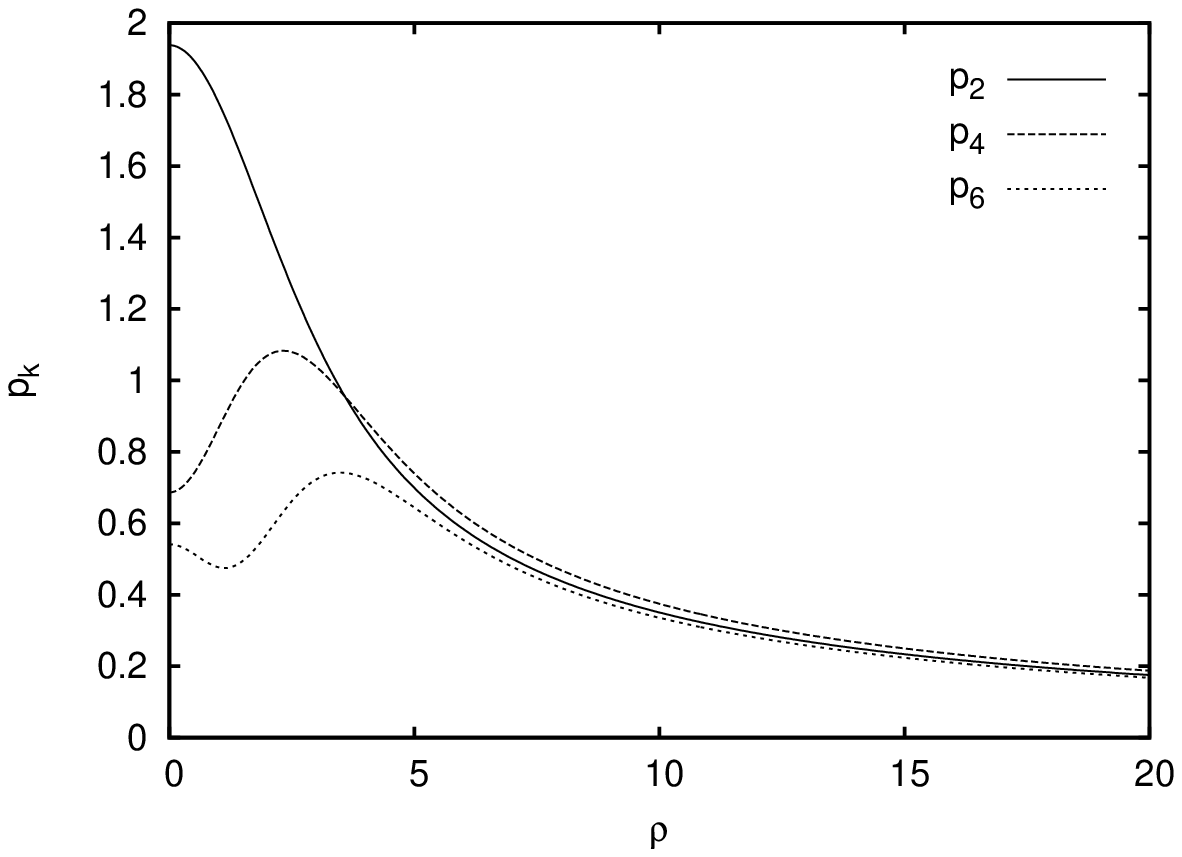,width=9.0cm} {The
  $p_k$ functions for the free scalar field case in $D=3$ dimensions.
\label{f:3dp}}
The obtained central values of $P_i$ and $p_i$ for
$i=2,4,6$ in $D=3,4,5$ are collected in Table \ref{tablepp}.
\TABULAR{|c|c|c|c|}{
\hline
  & $D=3$  & $D=4$  & $D=5$ \\
\hline
$P_{2c}$ & 2.04299 & 7.08429 & 28.0399 \\
$p_{2c}$ & 1.93832 & 4.42976 & 14.90729 \\
$P_{4c}$ & 0.658158 & -5.93174 & -348.868 \\
$p_{4c}$ & 0.686532 & -4.08270 & -200.353 \\
$P_{6c}$ & 0.557141 & 27.3950 & 9532.72 \\
$p_{6c}$ & 0.541339 & 17.9090 & 5500.18 \\
\hline}{
  Central values of the first three functions $P_i$ and $p_i$
  for the free scalar field in $3$, $4$ and $5$ spatial dimensions.
\label{tablepp}}
The chosen central values make all functions $P_i$ and $p_i$, and
consequently $\Phi_i$ and $\varphi_i$, tend to zero for $\rho\to\infty$.
Although for $i\geq 4$ $P_i$ and $p_i$ are not monotonically
decreasing functions, their central values represent well the
magnitude of these functions. Generally, the validity domain of an
asymptotic series ends where a higher order term starts giving larger
contributions than previous order terms. For $D=3$ the sixth order
$\varepsilon$ expansion can be expected to be valid even for as large
parameter values as $\varepsilon=1$. For $D=4$ this domain is
$\varepsilon<0.7$, while for $D=5$ it decreases to $\varepsilon<0.22$.

\subsection{$\varepsilon^4$ order for $D=6$} \label{sec:dsix}

As we have already stated in Subsection \ref{sec:sn}, if $D=6$ then
$\omega_2=0$, $s=S$ and the explicit form of the solution of the SN
equations is given by (\ref{6dss}). Introducing the new variables $z$
and $Z$ by
\begin{equation}\label{zzeq}
P_4=\frac{1}{3}(2Z+z) \,, \qquad p_4+\omega_4=\frac{1}{3}(Z-z) \,,
\end{equation}
equations (\ref{P4eq}) and (\ref{p4eq}) decouple,
\begin{eqnarray}
\frac{d^2z}{d\rho^2}+\frac{5}{\rho}\,\frac{d z}{d\rho}
-sz+\frac{3}{4}\beta_1s^3&=&0 \,,\label{eqz}\\
\frac{d^2Z}{d\rho^2}+\frac{5}{\rho}\,\frac{d Z}{d\rho}
+2sZ-\frac{9}{4}\beta_2s^3&=&0 \,,\label{eqZ}
\end{eqnarray}
where the constants $\beta_1$ and $\beta_2$ are defined by the
coefficients in the potential as
\begin{eqnarray}
\beta_1&=&1+\frac{80}{9}g_2^2-8g_3 \,,\\
\beta_2&=&1-\frac{80}{27}g_2^2+\frac{8}{3}g_3 \,.
\end{eqnarray}
For a free scalar field with $U(\Phi)=\Phi^2/2$ we have
$\beta_1=\beta_2=1$.  The general regular solution of (\ref{eqz}) can
be written in terms of the (complex indexed) associated Legendre
function $P$ as
\begin{equation}
z=\frac{144\beta_1\alpha^4(14+6\alpha^2\rho^2+\alpha^4\rho^4)}
{13(1+\alpha^2\rho^2)^4}
+\frac{C_1}{\alpha^2\rho^2}
P_{(i\sqrt{23}-1)/2}^{2}\left(\frac{1-\alpha^2\rho^2}
{1+\alpha^2\rho^2}\right) \,,
\end{equation}
where $C_1$ is some constant. The limiting value at $\rho\to\infty$ is
$z_\infty=-C_1\cosh(\pi\sqrt{23}/2)/\pi\approx -297.495\,C_1$. The
regular solution of (\ref{eqZ}) is
\begin{eqnarray}
Z&=&\frac{3888\beta_2\alpha^4(1-\alpha^2\rho^2)}
{7(1+\alpha^2\rho^2)^3}\ln(1+\alpha^2\rho^2)\\
&&+\frac{324\beta_2\alpha^6\rho^2(220+100\alpha^2\rho^2
-16\alpha^4\rho^4-\alpha^6\rho^6)}{35(1+\alpha^2\rho^2)^4}
+C_2\frac{\alpha^2\rho^2-1}
{(\alpha^2\rho^2+1)^3} \,,\nonumber
\end{eqnarray}
The limiting value at $\rho\to\infty$ is
$Z_\infty=-324\beta_2\alpha^4/35$, independently of $C_2$. Since $P_4$
must tend to zero, according to (\ref{zzeq}),
$z_\infty=648\beta_2\alpha^4/35$, fixing the constant $C_1$. Since the
mass of the field $\Phi$ is intended to remain $m=1$, the limit of
$p_4$ also has to vanish, giving
\begin{equation}
\omega_4=-\frac{324}{35}\beta_2\alpha^4 \,. \label{eqom4}
\end{equation}
This expression is not enough to fix $\omega_4$ yet, since $\alpha$ is
a free parameter. If $\beta_2>0$ then it is reasonable to use
(\ref{eqom4}) to set $\omega_4=-1$, thereby fixing the free parameter
$\alpha$ in the $\varepsilon^2$ order component of $\Phi$ and
$\varphi$. The change of the so far undetermined constant $C_2$
corresponds to a small rescaling of the parameter $\alpha$ in the
expression (\ref{6dss}). Its concrete value will fix the coefficient
$\omega_6$ in the expansion of the frequency.  The homogeneous parts
of the differential equations at higher $\varepsilon$ order will have
the same structure as those for $P_4$ and $p_4$.  Choosing the
appropriate homogeneous solutions all higher $\omega_k$ components can
be set to zero, yielding
\begin{equation}
\omega=\sqrt{1-\varepsilon^4} \ \  {\rm for}
 \ \  D=6  \ \ {\rm if} \ \ \beta_2>0 \,. \label{eq:om6d}
\end{equation}
This expression is valid for the free scalar field case with potential
$U(\Phi)=\Phi^2/2$ in $D=6$, since then $\beta_2=1$. For certain
potentials $\beta_2<0$, and one can use (\ref{eqom4}) to set
$\omega_4=1$. This case is quite unusual in the sense that the
frequency of the oscillon state is above the fundamental frequency
$m=1$. In the very special case, when $\beta_2=0$ the frequency
differs from the fundamental frequency only in $\varepsilon^6$ or
possibly higher order terms.

\subsection{Total energy and dilaton charge of oscillons}\label{ssec:energy}

Substituting (\ref{Phires}) and (\ref{varphires}) into the expression
(\ref{e:energysph}) of the total energy, we get
\begin{equation}
E = \varepsilon^{4-D}E_0
+\varepsilon^{6-D}E_1
+\mathcal{O}(\varepsilon^{8-D}) \,, 
 \label{e:energyeps}
\end{equation}
where
\begin{equation}
E_0=\frac{\pi^{D/2}}{\Gamma(D/2)}
\int_0^\infty d\rho\,\rho^{D-1}P_2^2 \,, \qquad
E_1=\frac{\pi^{D/2}}{\Gamma(D/2)}
\int_0^\infty d\rho\,\rho^{D-1}P_2(2P_4-P_2) \,. \label{e:energye0e1}
\end{equation}
Since $P_2=2S$, for $2<D<6$ we can use (\ref{e:s0s1}) and
(\ref{e:dmu}) to get
\begin{equation}
E_0=\frac{4\pi^{D/2}}{\Gamma(D/2)}(D-2)s_1\,. \label{e:energyeps2}
\end{equation}
The numerical values of $s_1$, $E_0$ and $E_1$ for $D=3,4,5$ are
listed in Table \ref{tablec1}.  
\TABULAR{|c|c|c|c|} {\hline
  & $D=3$  & $D=4$  & $D=5$ \\
  \hline
  $s_1$ & 3.50533 & 7.69489 & 10.4038 \\
  $E_0$ & 88.0985 & 607.565 & 1642.91 \\
  $E_1$ & 123.576 & 2522.10 & 31374.2 \\
  \hline} {The numerical values of $s_1$, $E_0$ and $E_1$ in $3$, 
  $4$ and $5$ spatial dimensions. \label{tablec1}} 
To the calculated order, i.e.\ up to $\varepsilon^{6-D}$, for $D=3$
and $D=4$ the energy is a monotonically increasing function of
$\varepsilon$, while for $D=5$ there is an energy minimum at
$\varepsilon=0.2288$. This result can only be taken as an estimate, as
the validity domain of an asymptotic series ends when two subsequent
terms are approximately equal.

For $D=6$ the leading order term in the total energy is
\begin{equation}
E=\frac{192\pi^3}{\alpha^2\varepsilon^2} \,. \label{e:energyd6}
\end{equation}
As we have already noted, for $D>6$ there are no finite energy
solutions.

The leading order $\varepsilon$ dependence of the dilaton charge for
$2<D<6$ is given by
\begin{equation}
Q=s_1\varepsilon^{4-D} \,,
\end{equation}
where we used the definition (\ref{eq:charge}), (\ref{e:s0s1}) and the
relation $\rho=\varepsilon r$. The dilaton charge for the $D=6$
oscillon is infinite. In higher orders in $\varepsilon$ the
proportionality between the dilaton charge and energy is violated.

\section{Time evolution of oscillons}

In this section we employ a numerical time evolution code in order to
simulate the actual behaviour of oscillons in the scalar-dilaton
theory. We use a fourth order method of line code with spatial
compactification in order to investigate spherically symmetric fields
\cite{FR2}. Our aim is to find configurations which are as closely
periodic as possible. To achieve this, we use initial data obtained
from the leading $\varepsilon^2$ terms of the small amplitude
expansion (\ref{Phires}) and (\ref{varphires}). The smaller the chosen
$\varepsilon$ is, the more closely periodic the resulting oscillating
state becomes. However, for moderate values of $\varepsilon$, it is
possible to improve the initial data by simply multiplying it by some
overall factor very close to $1$.

The main characteristics of the evolution of small amplitude initial
data depend on the number of spatial dimensions $D$. For $D=3$ and
$D=4$ oscillons appear to be stable.  If there is some moderate error
in the initial data, it will still evolve into an extremely long
living oscillating configuration, but its amplitude and frequency will
oscillate with a low frequency modulation. We employ a fine-tuning
procedure to minimize this modulation by multiplying the initial data
with some empirical factor. For $D=5$ and $D=6$ small amplitude
oscillons are not stable, having a single decay mode. In this case we
can use the fine-tuning method to suppress this decay mode, and make
long living oscillon states with well defined amplitude and
frequency. Without tuning in $D=5$ and $D=6$, in general, an initial
data evolves into a decaying state. The tuning becomes possible
because there are two possible ways of decay. One with a steady
outwards flux of energy, the other is through collapsing to a central
region first.

Having calculated several closely periodic oscillon configurations, it
is instructive to see how closely their total energy follow the
expressions (\ref{e:energyeps})-(\ref{e:energyd6}). Apart from
checking the consistency of the small amplitude and the time-evolution
approaches, this also gives information on how large $\varepsilon$
values the small amplitude expansion remains valid. The parameter
$\varepsilon$ for the evolving oscillon is calculated from the
numerically measured frequency by the expression
$\varepsilon=\sqrt{1-\omega^2}$. The results for $D=3$ are
presented on Fig.~\ref{f:en3d}.
\EPSFIGURE{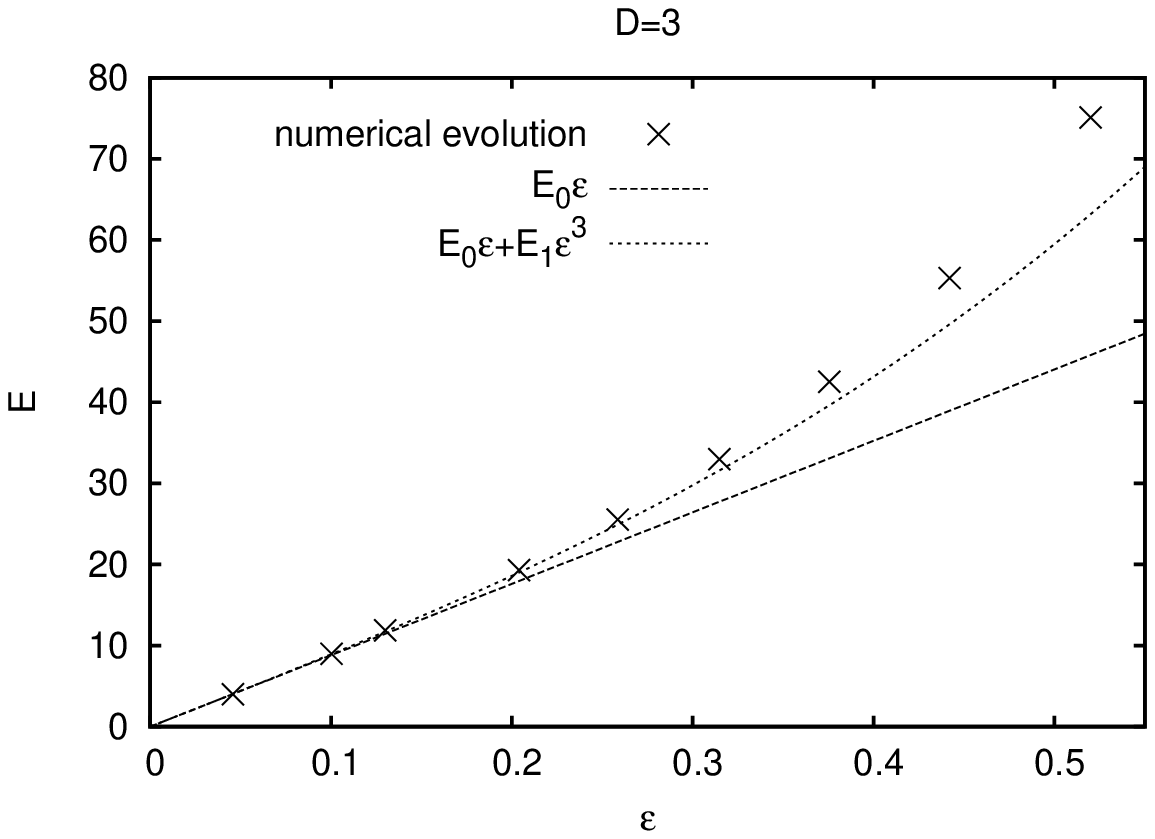,width=9.0cm} {Total energy
  of three-dimensional oscillons as a function of the parameter
  $\varepsilon$.
\label{f:en3d}}
In contrast to general relativistic oscillatons, there is no maximum
on the energy curve. This indicates that all three dimensional
oscillons in the dilaton theory are stable.  

The $\varepsilon$ dependence of the energy for $D=5$ is presented on
Fig.~\ref{f:en5d}. 
\EPSFIGURE{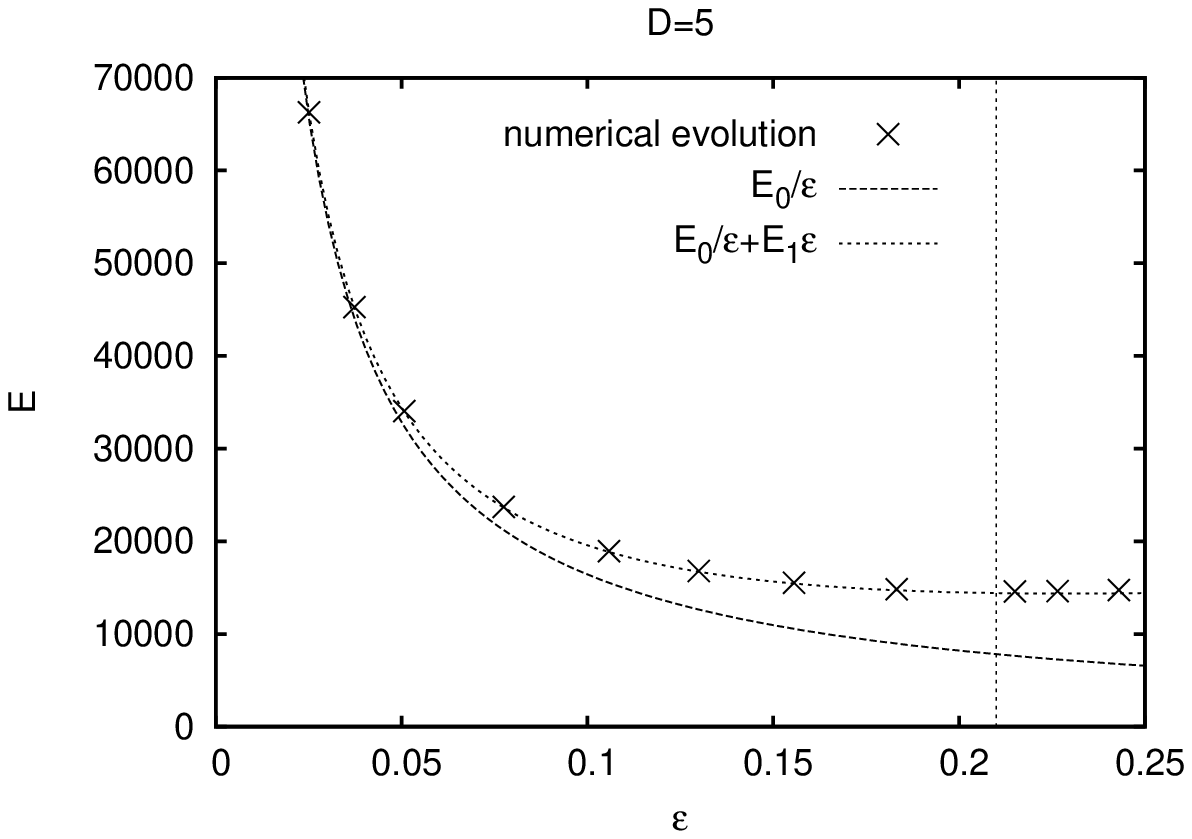,width=9.0cm} {Total energy of
  five-dimensional oscillons as a function of the parameter
  $\varepsilon$. The vertical line at $\varepsilon=0.21$ shows the
  place of the energy minimum. States to the right of it are stable,
  while those to the left have a single decay mode.
  \label{f:en5d}}
There is an energy minimum of the numerically
obtained states, approximately at $\varepsilon=0.21$, above which
oscillons are stable. The place of the minimum agrees quite well with
the value $\varepsilon=0.2288$ calculated in Subsection
\ref{ssec:energy} using the first two terms of the small amplitude
expansion. The behaviour of the energy close to the minimum is shown on 
Fig.~\ref{f:en5dmin}.
\EPSFIGURE{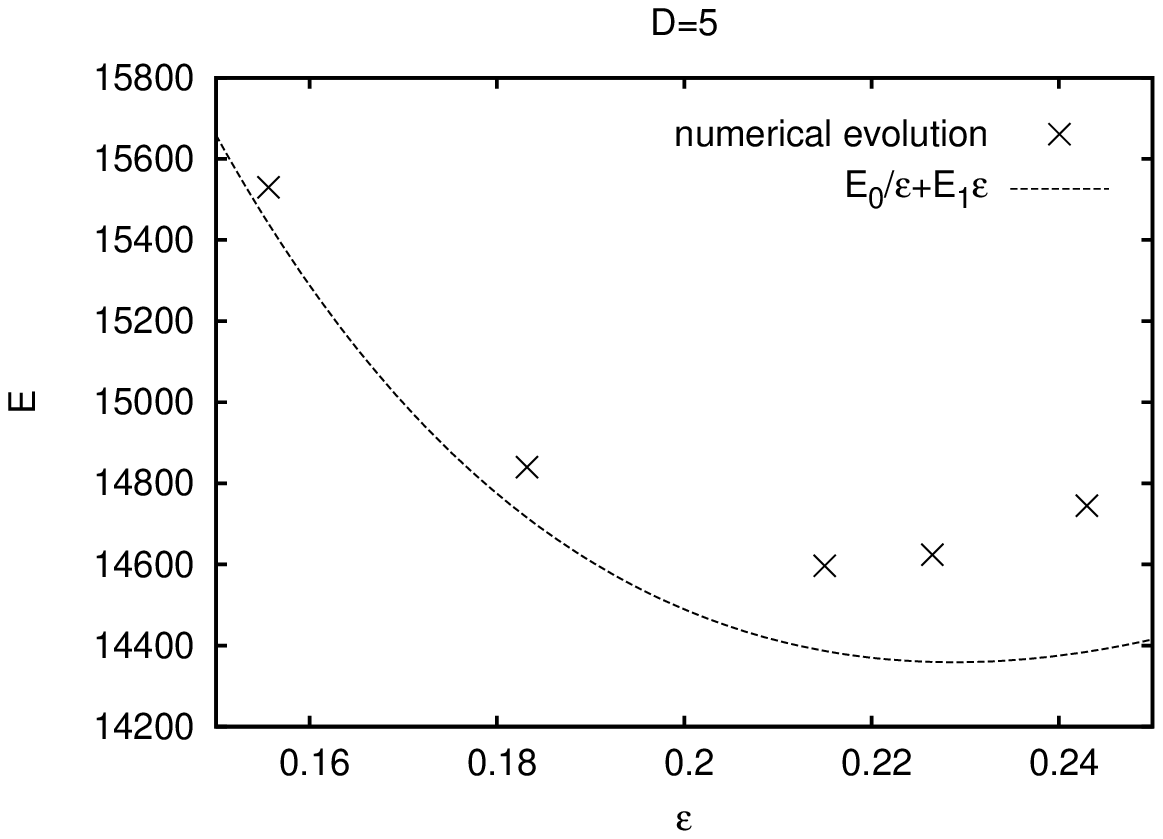,width=9.0cm} {The region of Fig.~\ref{f:en5d} 
near the energy minimum.  \label{f:en5dmin}}

We have also constructed oscillon states for $D=6$ dimensions. These
oscillons have quite large energy, due to the slow spatial decay of
the functions $\Phi$ and $\varphi$. For free massive scalar fields,
oscillons have frequency given by (\ref{eq:om6d}), i.e.\ an initial
data with a given $\varepsilon$ value will evolve to an oscillon state
with frequency approximately following
$\omega=\sqrt{1-\varepsilon^4}$. However, there are potentials, for
which the oscillation frequency is above the threshold $\omega=1$. For
example, this happens for the potential $U(\Phi)=\Phi^2(\Phi-2)^2/8$
with the choice $\kappa=1/2$.

In conclusion, in the dilaton-scalar theory oscillons follow the
stability pattern observed in the case of self-interacting scalar and
Einstein-Klein-Gordon theory; if $\varepsilon$ decreases with
decreasing energy, oscillons are stable, while if $\varepsilon$
increases with decreasing energy, oscillons are unstable. In other
words if the time evolution (i.e. energy loss) of an oscillon leads to
spreading of the core, the oscillon is stable, while oscillons are
unstable, if they have to contract with time evolution. The decreasing
or increasing nature of the energy, and hence empirically the
stability of the oscillating configurations, is well described by the
first two terms of the small amplitude expansion
(\ref{e:energyeps}). The result following from Eq.~(\ref{e:energyeps})
shows the existence of an energy minimum for $D>4$. This provides an
analytical argument for the existence of at least one unstable mode.
In particular, for $D=5$ spatial dimensions the frequency separating
the stable and unstable domains is determined by the small amplitude
expansion to satisfactory precision.

In order to study the instability in more detail numerically, we
compared the evolution of two almost identical initial data obtained
from the small amplitude expansion with $\varepsilon=0.05$. In order
to make the unstable state long living, a fine tuning procedure is
applied, multiplying the amplitude of the initial data by a factor
with value close to $1.0178$. The multiplicative factors used in the
two chosen initial values differ by $2.2\times 10^{-16}$. One of the
two initial data develops into a configuration decaying with a uniform
outward current of energy, the other through collapsing to a high
density state first. On Fig.\ \ref{f:phidiff} the time evolution of
the difference of the central value of the dilaton fields in the two
states $\Delta\varphi=\varphi_1-\varphi_2$ is shown.
\EPSFIGURE{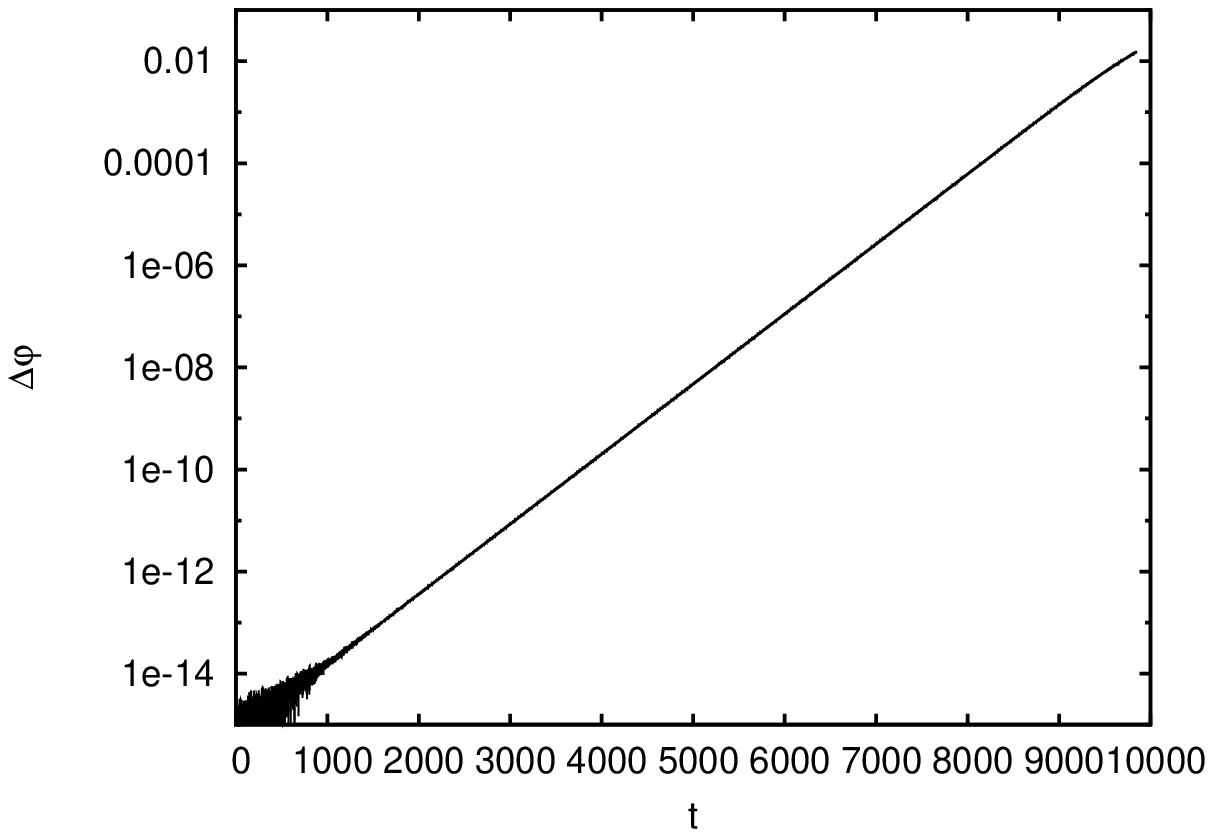,width=9.0cm} {Increase in the difference of the 
dilaton field $\varphi$ for two similar configurations. 
 \label{f:phidiff}}
The curve follows extremely well the exponential increase described by
\begin{equation}
\Delta\varphi=6.583\times 10^{-16}\exp(0.003157\,t) \,,
\end{equation}
showing that there is a single decay mode growing exponentially. 
The difference of the scalar fields, $\Delta\Phi=\Phi_1-\Phi_2$, 
grows with the same exponent. The spatial dependence of the decaying 
mode is illustrated on Fig.\ \ref{f:ppdiff}, where $\Delta\Phi$ is 
plotted at several moments of time corresponding to the maximum of 
$\Phi_1$ at the center. 
\EPSFIGURE{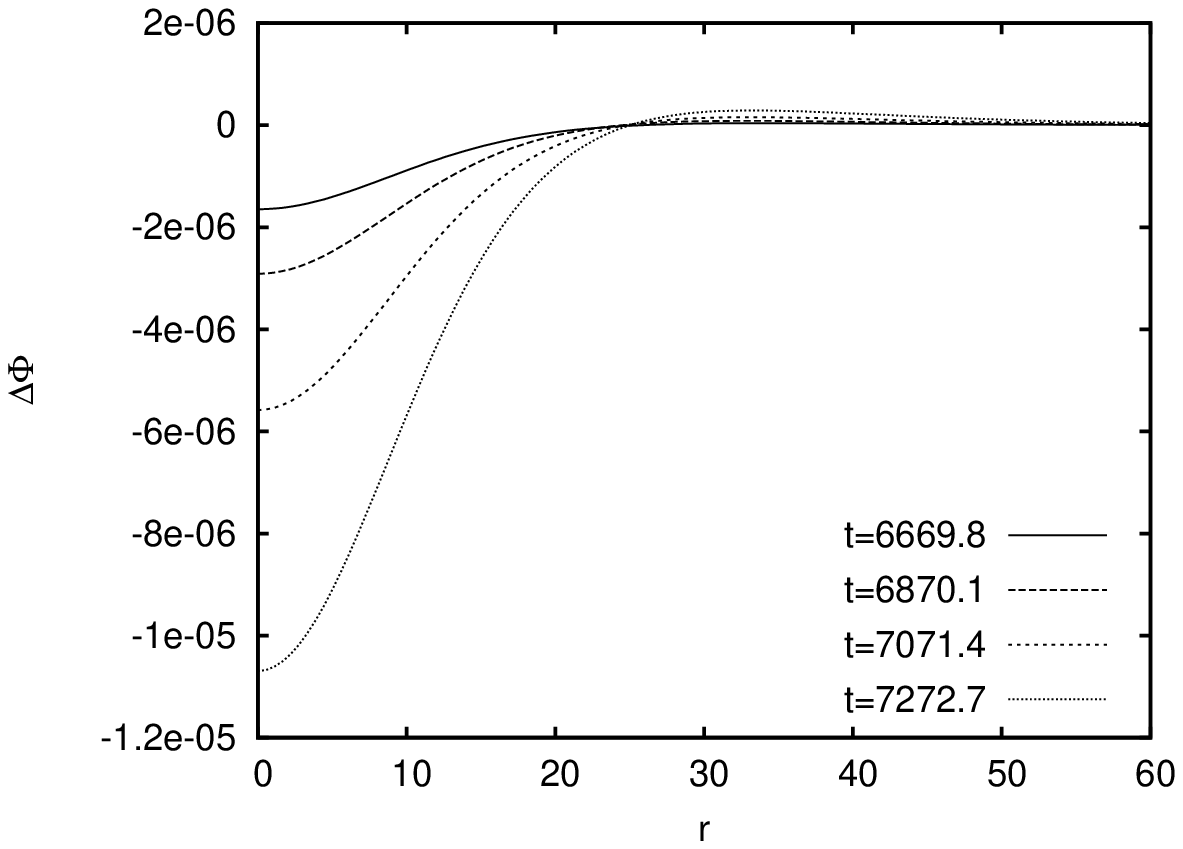,width=9.0cm} {Radial behaviour of the difference
of the scalar fields of two very similar configurations. At the chosen
moments of time the scalar is maximal at the center, and the
subsequent moments are separated by $32$ oscillations. 
\label{f:ppdiff}}

Our numerical results strongly indicate that oscillons in the
scalar-dilaton theory are unstable for $D>4$, and they admit a single
decay mode. For the single scalar field system the instability arises
for $D>2$ (see Ref.\ \cite{moredim}), but the decay modes have been
calculated analytically only in some very special cases. The scalar
theory with potential $U(\phi)=\phi^2(1-\ln\phi^2)$ admits exactly
time-periodic breathers in any dimensions. The stability of these
breathers in three dimensions has been investigated in detail in
\cite{koutv1}\cite{koutv2}. It has been found that these breathers
always admit a single unstable mode. It needs further studies whether
an analysis along the lines of Ref.~\cite{koutv1} can also be applied
to more general potentials in the small amplitude limit, and whether
it can be generalized to the case when the scalar is coupled to a
dilaton field.

\section{Determination of the energy loss rate}\label{s:borel}

Although oscillons are extremely long living, generally they are not
exactly periodic.  In this section we calculate how the energy loss
rate depends on the oscillon frequency for small amplitude
configurations. To simplify the expressions in this section we
consider a massive free scalar field, i.e.\ 
$U(\Phi)=m^2\Phi^2/2$. We assume that $2<D<6$, since then the scalar
field tends to zero exponentially for large $\rho$.

The outgoing radiation will dominantly be in the dilaton field and the
radiation amplitude will have the $\varepsilon$ dependence:
$\varepsilon\,\exp\left(-2Q_D/\varepsilon\right)$. In
Refs. \cite{FFHM} and \cite{moredim} we have used two different methods
for determining the $\varepsilon$ independent part of the radiation
amplitude: Borel summation and solution of the complexified mode
equations numerically. In this paper we will use an analytic method based on Borel-summing the asymptotic
series in the neighborhood of its singularity in the complex plane. Other potentials which are symmetric
around their minima can be treated analogously. If the potential is
asymmetric only the numerical method could be used. This phenomenon is
in complete analogy with the problem arising with a single scalar
field considered in Ref. \cite{FFHM}.

\subsection{Singularity of the small $\varepsilon$ expansion}

We first investigate the complex extension of the functions obtained
by the small amplitude expansion in Sec.~\ref{sec:smallampl}.
Extending the solutions $s$ and $S$ of the Schr\"odinger-Newton
equations (\ref{Seq}) and (\ref{seq}) to complex $\rho$ coordinates
they both have pole singularities on the imaginary axis of the complex
plane. We consider the closest pair of singularities to the real axis,
since these will give the dominant contribution to the energy loss.
They are located at $\rho=\pm iQ_D$. The numerically calculated values
of $Q_D$ are listed in Table \ref{tableqd} for spatial dimensions
$D=3,4,5$.
\TABULAR{|c|c|}
{\hline
$\ \ \ D \ \ \ $  & $\ \ \ \ \ Q_{D} \ \ \ \ \ $ \\
\hline
3 & 3.97736\\
4 & 2.30468\\
5 & 1.23595\\
\hline}
{The distance $Q_D$ between the real axis and the pole of
  the fundamental solution of the SN equation for various spatial
  dimensions $D$. \label{tableqd}}
Let us measure distances from the upper singularity by a coordinate
$R$ defined as
\begin{equation}
\rho=iQ_D+R\,.	
\end{equation}
Close to the pole we can expand the SN equations, and obtain that $s$
and $S$ have essentially the same behaviour,
\begin{equation}
s=\pm S=-\frac{6}{R^2}-\frac{6i(D-1)}{5Q_DR}
-\frac{(D-1)(D-51)}{50Q_D^2}+{\cal O}(R) \ , \label{eq:sexp}
\end{equation}
even though they clearly differ on the real axis.  Since for symmetric
potentials we can always substitute $\Phi$ by $-\Phi$, we choose the
positive sign for $S$ in (\ref{eq:sexp}). This choice is compatible
with the sign of $S$ used on the real axis at the small amplitude
expansion section. We note that for $D>1$ there are logarithmic terms
in the expansion of $s$ and $S$, starting with terms proportional to
$R^4\ln R$.  According to (\ref{eq:defss}), the functions determining
the leading $\varepsilon^2$ parts of $\varphi$ and $\Phi$ in this case
are
\begin{equation}
p_2=s+1 \,, \qquad P_2=2S \,.\label{eq:p2p2}
\end{equation}
Substituting these into the equations (\ref{P4eq}) and (\ref{p4eq}),
the $\varepsilon^4$ order contributions $p_4$ and $P_4$ can also be
expanded around the pole
\begin{eqnarray}
p_4&=&-\frac{1161}{52R^4}+\frac{324i(D-1)\ln R}{35Q_DR^3}
+\frac{c_{-3}}{R^3}+{\cal O}\left(\frac{\ln R}{R^2}\right)
\label{eq:p4pole}\\
P_4-2p_4&=&\frac{81}{13R^4}
+\frac{18i(D-1)}{5Q_DR^3}
+{\cal O}\left(\frac{1}{R^2}\right) \,,\label{eq:P4pole}
\end{eqnarray}
where the constant $c_{-3}$ can only be determined from the specific
behaviour of the functions on the real axis, namely from the
requirement of the exponential decay of $P_4$ for large real $\rho$.

\subsection{Fourier mode expansion}

Since all terms of the small amplitude expansion (\ref{sumphi}) are
asymptotically decaying, i.e.\ localized functions, the small
amplitude expansion can be successfully applied to the core region of
oscillons. However it cannot describe the exponentially small
radiative tail responsible for the energy loss. Instead of studying a
slowly varying frequency radiating oscillon configuration it is
simpler to consider exactly periodic solutions having a large core and
a very small amplitude standing wave tail. We look for periodic
solutions with frequency $\omega$ by Fourier expanding the scalar and
dilaton field as
\begin{equation}
\Phi=\sum_{k=0}^{N_F}\Psi_{k}\cos(k\omega t)
\,, \qquad
\varphi=\sum_{k=0}^{N_F}\psi_{k}\cos(k\omega t) \,. \label{eq:ppexp}
\end{equation}
Although, in principle, the Fourier truncation order $N_F$ should tend
to infinity, one can expect very good approximation for moderate
values of $N_F$.  In (\ref{eq:ppexp}) we denoted the Fourier
components by psi instead of phi to distinguish them from the small
$\varepsilon$ expansion components in (\ref{sumphi}). Since in this
section we only deal with an self-interaction free scalar field with a
trivially symmetric potential,
\begin{equation}
\Psi_{2k}=0 \,, \qquad \psi_{2k+1}=0 \,, \qquad
{\rm for\ integer} \ k \,.
\end{equation}
We note that the absence of sine terms in (\ref{eq:ppexp}) is
equivalent to the assumption of time reflexion symmetry at $t=0$. This
assumption appears reasonable physically, and we have seen in
Sec.~\ref{sec:smallampl} that it holds in the small amplitude
expansion framework.

For small amplitude configurations we can establish the connection
between the expansions (\ref{sumphi}) and (\ref{eq:ppexp}) by
comparing to (\ref{Phires}) and (\ref{varphires}), obtaining
\begin{eqnarray}
\Psi_1&=&\varepsilon^2P_2
+\varepsilon^4P_4+\varepsilon^6P_6+{\cal O}(\varepsilon^8)
\ , \label{eq:Psifour}\\
\Psi_3&=&\varepsilon^6\frac{P_2^3}{256}+{\cal O}(\varepsilon^8) \ ,
\\
\psi_0&=&\varepsilon^2p_2
+\varepsilon^4p_4+\varepsilon^6p_6+{\cal O}(\varepsilon^8) \ , \\
\psi_2&=&-\varepsilon^4\frac{P_2^2}{16}
-\varepsilon^6
\frac{1}{32}\left[4P_2P_4-(p_2-1)P_2^2-
\left(\frac{d P_2}{dR}\right)^2\right]
+{\cal O}(\varepsilon^8) \ .\label{eq:psifour}
\end{eqnarray}

Let us define a coordinate $y$ for an ``inner region'' by
$R=\varepsilon y$. This coordinate will have the same scale as the
original radial coordinate $r$, since they are related as
\begin{equation}
r=\frac{iQ_D}{\varepsilon}+y\,.	
\end{equation}
The ``inner region'' $\vert R \vert \ll 1$ is not small in the $y$
coordinate; if $\varepsilon\to0$ then $\varepsilon \vert y\vert=\vert
R\vert\ll 1$ but $\vert y\vert\to\infty$.  Using the coordinate $y$
and substituting (\ref{eq:sexp})-(\ref{eq:P4pole}) into the small
amplitude Fourier mode expressions
(\ref{eq:Psifour})-(\ref{eq:psifour}), we obtain that the leading
asymptotic behaviour of the Fourier modes for $\vert y\vert\to\infty$
can be written as
\begin{eqnarray}
\Psi_1&=&-\frac{12}{y^2}-\frac{999}{26y^4}
+\varepsilon\ln\varepsilon\frac{648i(D-1)}{35Q_Dy^3}\nonumber\\
&&+\varepsilon\left[\frac{6i(D-1)}{5Q_Dy}\left(\frac{3}{y^2}
+\frac{108\ln y}{7y^2}-2\right)
+\frac{2c_{-3}}{y^3}\right] +\ldots \ , \label{eq:Psi1ser}\\
\Psi_3&=&-\frac{27}{4y^6}
-\varepsilon\frac{81i(D-1)}{20Q_Dy^5}+\ldots \ ,  \\
\psi_0&=&-\frac{6}{y^2}-\frac{1161}{52y^4}
+\varepsilon\ln\varepsilon\frac{324i(D-1)}{35Q_Dy^3}\nonumber\\
&&+\varepsilon\left[\frac{6i(D-1)}{5Q_Dy}\left(
\frac{54\ln y}{7y^2}-1\right)
+\frac{c_{-3}}{y^3}\right] +\ldots \ ,  \\
\psi_2&=&-\frac{9}{y^4}
-\varepsilon\frac{18i(D-1)}{5Q_Dy^3}+\ldots \ . \label{eq:psi2ser}
\end{eqnarray}
These expressions are simultaneous series in $1/y$ and in
$\varepsilon$.

\subsection{Fourier mode equations}

In order to obtain finite number of Fourier mode equations with finite
number of terms, when substituting (\ref{eq:ppexp}) into the field
equations (\ref{evolPhi}) and (\ref{evolvarphi}) we Taylor expand and
truncate the exponential
\begin{equation}
e^{-\varphi}=\sum_{k=0}^{N_e}\frac{1}{k!}(-\varphi)^k \,.	
\end{equation}
We need to carefully check how large $N_e$ should be chosen to have
only a negligible influence to the calculated results. For $n\leq N_F$
the Fourier mode equations have the form
\begin{eqnarray}
\left(\frac{\mathrm{d^2}}{\mathrm{d}r^2}
+\frac{D-1}{r}\,\frac{\mathrm{d}}{\mathrm{d}r}
+n^2\omega^2-1\right)\Psi_n&=&F_n \ , \label{eq:Psin}\\
\left(\frac{\mathrm{d^2}}{\mathrm{d}r^2}
+\frac{D-1}{r}\,\frac{\mathrm{d}}{\mathrm{d}r}
+n^2\omega^2\right)\psi_n&=&f_n \ ,\label{eq:psin}
\end{eqnarray}
where we have collected the nonlinear terms to the right hand sides,
and denoted them with $F_n$ and $f_n$. These are polynomial
expressions involving various $\Psi_k$ and $\psi_k$, with quickly
increasing complexity when increasing the truncation orders $N_F$ and
$N_e$. The solution of (\ref{eq:Psin}) and (\ref{eq:psin}) yields the
intended quasibreathers, with a localized core and a very small
amplitude oscillating tail. For small amplitude configurations the
functions $\Psi_k$ and $\psi_k$ will have poles at the complex $r$
plane, just as we have seen in the small amplitude expansion
formalism. In order to calculate the tail amplitude it is necessary to
investigate the Fourier mode equations instead of the equations
obtained in Sec.~\ref{sec:smallampl}. Although in the Fourier
decomposition method we have not defined a small amplitude parameter
yet, motivated by (\ref{eq:omeps}), we can, in general, define
$\varepsilon$ as
\begin{equation}
\varepsilon=\sqrt{1-\omega^2} \,.
\end{equation}
Dropping $\mathcal{O}(\varepsilon^2)$ terms, in the neighborhood of
the singularity the mode equations take the form
\begin{eqnarray}
\left(\frac{\mathrm{d^2}}{\mathrm{d}y^2}
+\varepsilon\frac{D-1}{iQ_D}\,\frac{\mathrm{d}}{\mathrm{d}y}
+n^2-1\right)\Psi_n&=&F_n \ , \label{eq:Psin2}\\
\left(\frac{\mathrm{d^2}}{\mathrm{d}y^2}
+\varepsilon\frac{D-1}{iQ_D}\,\frac{\mathrm{d}}{\mathrm{d}y}
+n^2\right)\psi_n&=&f_n \ .\label{eq:psin2}
\end{eqnarray}
We look for solutions of these equations that satisfy
(\ref{eq:Psi1ser})-(\ref{eq:psi2ser}) as boundary conditions for
$\vert y\vert \to\infty$ for $-\pi/2<\arg y<0$. This corresponds to
the requirement that the functions decay to zero without any
oscillating tails for large $r$ on the real axis. The small correction
corresponding to the nonperturbative tail of the quasibreather will
arise in the imaginary part of the functions on the $\textrm{Re}\,y=0$
axis.

\subsection{$\varepsilon\to 0$ limit near the pole}

For very small $\varepsilon$ values one can neglect the terms
proportional $\varepsilon$ on the left hand sides of (\ref{eq:Psin2})
and (\ref{eq:psin2}). In this limit the there is no dependence on the
number of spatial dimensions $D$. We investigate this simpler system
first, and consider finite but small $\varepsilon$ corrections later
as perturbations to it. We expand the solution of (\ref{eq:Psin2}) and
(\ref{eq:psin2}) (with $\varepsilon=0$) in even powers of $1/y$,
\begin{equation}
\Psi_{2k+1}=\sum_{j=k+1}^{\infty}A_{2k+1}^{(j)}\frac{1}{y^{2j}}
\,, \qquad
\psi_{2k}=\sum_{j=k+1}^{\infty}a_{2k}^{(j)}\frac{1}{y^{2j}} \,.
\label{eq:ppy}
\end{equation}
We illustrate our method by a minimal system where radiation loss can
be studied, namely the case with $N_F=3$ and $N_e=1$. Then the mode
equations are still short enough to print:
\begin{eqnarray}
\frac{d^2\psi_0}{d y^2}&=&\frac{1}{4}(\psi_0-1)(\Psi_1^2+\Psi_3^2)
+\frac{1}{8}\psi_2\Psi_1(\Psi_1+2\Psi_3) \,, \label{eq:mins1}\\
\frac{d^2\Psi_1}{d y^2}&=&-\psi_0\Psi_1
-\frac{1}{2}\psi_2(\Psi_1+\Psi_3) \,,\\
\frac{d^2\psi_2}{d y^2}+4\psi_2&=&
\frac{1}{4}\Psi_1(\psi_0-1)(\Psi_1+2\Psi_3)
+\frac{1}{4}\psi_2(\Psi_1^2+\Psi_1\Psi_3+\Psi_3^2) \,,\\
\frac{d^2\Psi_3}{d y^2}+8\Psi_3&=&-\frac12 \psi_2\Psi_1-\psi_0\Psi_3
\,.\label{eq:mins4}
\end{eqnarray}
When looking for solution of these equations in the form of the
$1/y^2$ expansion (\ref{eq:ppy}), only one ambiguity arises, the sign
of $A_0^{(1)}$. Choosing it to be negative, the first few terms of the
expansion turn out to be
\begin{eqnarray}
\psi_0&=&-\frac{6}{y^2}-\frac{837}{52y^4}
+\mathcal{O}\left(\frac{1}{y^6}\right) \,, \\
\Psi_1&=&-\frac{12}{y^2}-\frac{459}{26y^4}
+\mathcal{O}\left(\frac{1}{y^6}\right) \,, \\
\psi_2&=&-\frac{9}{y^4}-\frac{1845}{52y^6}
+\mathcal{O}\left(\frac{1}{y^8}\right) \,, \\
\Psi_3&=&-\frac{27}{4y^6}-\frac{2565}{416y^8}
+\mathcal{O}\left(\frac{1}{y^{10}}\right) \,.
\end{eqnarray}
The first terms agree with those of
(\ref{eq:Psi1ser})-(\ref{eq:psi2ser}) obtained by the small amplitude
expansion. The difference in the $1/y^4$ terms of $\psi_0$ and
$\Psi_1$ are caused by the too low truncation for the Taylor expansion
of the exponential. For $N_e\geq 2$ these terms agree as well.

When increasing $N_F$ and $N_e$ growing number of additional terms
appear on the right hand sides of (\ref{eq:mins1})-(\ref{eq:mins4}),
and the number of mode equations rise to $N_F+1$. These complicated
mode equations can be calculated and $1/y^2$ expanded using an
algebraic manipulation program. However, apart from a factor, the
leading order behaviour of the coefficients $a_{2}^{(n)}$ and
$A_{3}^{(n)}$ for large $n$ will remain the same as that of the
minimal system (\ref{eq:mins1})-(\ref{eq:mins4}). The large $n$
behaviour of these coefficients will be essential for the calculation
of the nonperturbative effects resulting in radiation loss for
oscillons.

Starting from the free system, consisting of the linear terms on the
left hand sides, it is easy to see that the mode equations are
consistent with the following asymptotic (large $n$) behavior of the
coefficients,
\begin{eqnarray}
a_2^{(n)}&\sim& k\,(-1)^n\,\frac{(2n-1)!}{4^n}\label{eq:asym1a}\\
a_0^{(n)},\, A_1^{(n)},\, A_3^{(n)} \, &\ll& a_2^{(n)}\,,\label{eq:asym1b}
\end{eqnarray}
where $k$ is some constant. The value of $k$ can be obtained to a
satisfactory precision by substituting the $1/y$ expansion into the
mode equations and explicitly calculating the coefficients to up to
high orders in $n$. In practice, using an algebraic manipulation
software, we have calculated coefficients up to order $n=50$. The
dependence of $k$ on the order of the Fourier expansion is given in
Table \ref{table:k}.  \TABULAR{|c|c|c|} {\hline
  $N_F$  & $N_e^{(min)}$ & $k$ \\
  \hline
  $3$ & $5$  & $3.71\times10^{-3}$\\
  $4$ & $7$  & $3.12\times10^{-6}$\\
  $5$ & $8$  & $6.03\times10^{-9}$\\
  $6$ & $10$ & $4.61\times10^{-13}$\\
  \hline} {Dependence of the constant $k$ on the considered Fourier
  components $N_F$. The second column lists the minimal exponential
  expansion order $N_e$ which is necessary to get the $k$ value with
  the given precision.  \label{table:k}}
The results strongly indicate that in the $N_f,N_e\to \infty$ limit
$k=0$. We do not yet understand what is the deeper reason or symmetry
behind this. Hence, instead of (\ref{eq:asym1a}) and
(\ref{eq:asym1b}), the correct asymptotic behavior is
\begin{eqnarray}
A_3^{(n)}&\sim& K\,(-1)^n\,\frac{(2n-1)!}{8^n}\label{eq:a3asympt}\\
a_0^{(n)},\, A_1^{(n)},\, a_2^{(n)} \, &\ll& A_3^{(n)}\,.
\end{eqnarray}
Taking at least $N_F=6$ and $N_e=9$, the numerical value of the
constant turns out to be $K=-0.57\pm0.01$. The above results indicate
that the outgoing radiation is in the $\Psi_3$ scalar mode instead of
being in the $\psi_2$ dilaton mode. This conclusion is valid only in
the framework of the approximation employed in the present subsection,
i.e. when dropping the terms proportional to $\varepsilon$ in
(\ref{eq:Psin2}) and (\ref{eq:psin2}). As we will see in the next
subsection, the situation will change to be just the opposite when
taking into account $\varepsilon$ corrections.

All terms of the expansion (\ref{eq:ppy}) are real on the imaginary
axis $\mathrm{Re}\,y=0$. However, using the Borel-summation procedure
it is possible to calculate there an exponentially small correction to
the imaginary part. We will only sketch how the summation is done, for
details see \cite{FFHM} and \cite{Pomeau}. We illustrate the method by
applying it to $\Psi_3$. The first step is to define a Borel summed
series by
\begin{equation}
  V(z)=\sum_{n=2}^{\infty}\frac{A_3^{(n)}}{(2n)!}\;
  z^{2n}\sim\sum_{n=2}^{\infty}K\,\frac{(-1)^n}{2n}\;
  \left(\frac{z}{\sqrt 8}\right)^{2n}
=-\frac{K}{2}\ln\left(1+\frac{z^2}{8}\right)\;. \label{e:borelsum}
\end{equation}
This series has logarithmic singularities at $z=\pm i \sqrt 8$.  The
Laplace transform of $V(z)$ will give us the Borel summed series of
$\Psi_3(y)$ which we denote by $\widehat{\Psi}_3(y)$
\begin{equation}
\widehat{\Psi}_3(y)=\int_{0}^{\infty} \;\mathrm{d}t\,
e^{-t}V\left(\frac{t}{y}\right)\,.\label{borelint}
\end{equation}
The choice of integration contour corresponds to the requirement of exponential decay on the real axis.
 The logarithmic singularity of $V\left(t/y\right)$ does not contribute
to the integral and integrating on the branch cut starting from it
yields the imaginary part
\begin{equation}
{\rm Im}\,\widehat{\Psi}_3(y)=\int_{i \sqrt8 \,y}^{\infty}
\;\mathrm{d}t\; e^{-t}\,\frac{K\pi}{2}
=\frac{K\pi}{2}\,\exp\left(-i \sqrt8 \,y\right)\,.
\label{e:borel}
\end{equation}
A similar calculation for the $\psi_2$ dilaton mode yields
\begin{equation}
{\rm Im}\,\widehat{\psi}_2(y)
=\frac{k\pi}{2}\,\exp\left(-2iy\right)\,.
\end{equation}
Since $k=0$, this mode is vanishing now. However, as we will show in
the next subsection, when taking into account order $\varepsilon$
corrections a similar expression for $\psi_2$ with
$\exp\left(-2iy\right)$ behaviour arise, which, due to its slower
decay, will become dominant when $\mathrm{Im}\,y\to -\infty$.  The
continuation to the real axis of these imaginary corrections turns out
to be closely related to the asymptotically oscillating mode
responsible for the slow energy loss of oscillons.

\subsection{Order $\varepsilon$ corrections near the pole}

Before discussing the issue of matching the imaginary correction
calculated in the neighborhood of the singularity to the solution of
the field equation on the real axis we deal with the corrections
arising when taking into account the terms proportional to
$\varepsilon$ in the mode equations (\ref{eq:Psin2}) and
(\ref{eq:psin2}). We denote the solutions obtained in the previous
subsection by $\psi_n^{(0)}$ and $\Psi_n^{(0)}$, and linearize the
mode equations around them by defining
\begin{equation}
\psi_{n}=\psi_{n}^{(0)}+\widetilde{\psi}_{n} \,, \qquad
\Psi_{n}=\Psi_{n}^{(0)}+\widetilde{\Psi}_{n} \,.
\end{equation}
The mode equations take the form
\begin{eqnarray}
\left(\frac{{\rm d}^2}{{\rm d} y^2} +n^2\right) \widetilde{\psi}_n
+\varepsilon\frac{D-1}{iQ_D}\,\frac{{\rm d}\psi_n^{(0)}}{{\rm d}y}
&=&\sum_{m}\frac{\partial f_n}{\partial \psi_m}\widetilde{\psi}_m
+\sum_{m}\frac{\partial f_n}{\partial \Psi_m}\widetilde{\Psi}_m \,,
\label{eq:lin1a}\\
\left(\frac{{\rm d}^2}{{\rm d} y^2} +n^2-1\right) \widetilde{\Psi}_n
+\varepsilon\frac{D-1}{iQ_D}\,\frac{{\rm d}\Psi_n^{(0)}}{{\rm d}y}
&=&\sum_{m}\frac{\partial F_n}{\partial \psi_m}\widetilde{\psi}_m
+\sum_{m}\frac{\partial F_n}{\partial \Psi_m}\widetilde{\Psi}_m\,,
\label{eq:lin1b}
\end{eqnarray}
where the partial derivatives on the right hand sides are taken at
$\Psi_n=\Psi_n^{(0)}$ and $\psi_n=\psi_n^{(0)}$. The small dimensional
corrections $\widetilde{\psi}_n$ and $\widetilde{\Psi}_n$ have parts
of order both $\varepsilon \ln \varepsilon$ and $\varepsilon$.

The linearized equations (\ref{eq:lin1a}) and (\ref{eq:lin1b}) are
solved to $\varepsilon \ln \varepsilon$ order by the following
functions:
\begin{eqnarray}
\widetilde{\psi}_n&=&\varepsilon \ln \varepsilon\,C
\frac{{\rm d}\psi_n^{(0)}}{{\rm d}y} \,,\label{e:zeromode1}\\
\widetilde{\Psi}_n&=&\varepsilon \ln \varepsilon\,C
\frac{{\rm d}\Psi_n^{(0)}}{{\rm d}y}\,, \label{e:zeromode2}
\end{eqnarray}
where $C$ is an arbitrary constant. The reason for this is quite
simple: in $\varepsilon \ln \varepsilon$ order the terms proportional
to $\varepsilon$ on the left hand sides are negligible and we get the
$\varepsilon=0$ equation linearized about the original solution. Our
formula simply gives the zero mode of this equation. The constant $C$
is determined by the appropriate behaviour when continuing back our
functions to the real axis. This can be ensured by requiring agreement
with the first few terms of the small amplitude expansion formulae
(\ref{eq:Psi1ser})-(\ref{eq:psi2ser}), yielding
\begin{equation}
C=\frac{27i(D-1)}{35Q_D}\,. \label{eq:cc}
\end{equation}

In the small amplitude expansion (\ref{eq:Psi1ser})-(\ref{eq:psi2ser})
to every term of order $\varepsilon \ln \varepsilon$ corresponds a
term of order $\varepsilon$ which we get by changing $\ln \varepsilon$
to $\ln y$. Thus, we define the new variables $\overline{\psi}_n$ and
$\overline{\Psi}_n$ to describe the $\varepsilon$ order small
perturbations by
\begin{eqnarray}
\widetilde{\psi}_n&=&\varepsilon \ln \varepsilon\,C
\frac{{\rm d}\psi_n^{(0)}}{{\rm d}y}
+\varepsilon\left(C\ln y\,\frac{{\rm d}\psi_n^{(0)}}{{\rm d}y}
+\overline{\psi}_n\right) \,,\label{e:dimcor1a}\\
\widetilde{\Psi}_n&=&\varepsilon \ln \varepsilon\,C
\frac{{\rm d}\Psi_n^{(0)}}{{\rm d}y}
+\varepsilon\left(C\ln y\,\frac{{\rm d}\Psi_n^{(0)}}{{\rm d}y}
+\overline{\Psi}_n\right)\,. \label{e:dimcor1b}
\end{eqnarray}
Substituting into the linearized equations (\ref{eq:lin1a}) and
(\ref{eq:lin1b}) we see that all terms containing $\ln y$ cancel out,
\begin{eqnarray}
&&\left(\frac{{\rm d}^2}{{\rm d} y^2} +n^2 \right)\overline{\psi}_n
+\frac{C}{y^2}\left(2y\frac{{\rm d^2}\psi_n^{(0)}}{{\rm d}y^2}
-\frac{{\rm d}\psi_n^{(0)}}{{\rm d}y}\right)
+\frac{D-1}{iQ_D}\,\frac{{\rm d}\psi_n^{(0)}}{{\rm d}y}=\nonumber\\
&&\qquad=\sum_{m}\frac{\partial f_n}{\partial \psi_m}\overline{\psi}_m
+\sum_{m}\frac{\partial f_n}{\partial \Psi_m}\overline{\Psi}_m \,,
\label{e:epscor}\\
&&\left(\frac{{\rm d}^2}{{\rm d} y^2} +n^2 -1\right)\overline{\Psi}_n
+\frac{C}{y^2}\left(2y\frac{{\rm d^2}\Psi_n^{(0)}}{{\rm d}y^2}
-\frac{{\rm d}\Psi_n^{(0)}}{{\rm d}y}\right)
+\frac{D-1}{iQ_D}\,\frac{{\rm d}\Psi_n^{(0)}}{{\rm d}y}=\nonumber\\
&&\qquad=\sum_{m}\frac{\partial F_n}{\partial \psi_m}\overline{\psi}_m
+\sum_{m}\frac{\partial F_n}{\partial \Psi_m}\overline{\Psi}_m\,.
\label{e:epscor2}
\end{eqnarray}
If $C$ is given by (\ref{eq:cc}), $\overline{\psi}_n$ and
$\overline{\Psi}_n$ turn out to be algebraic asymptotic series
which are analytic in $y$. Let us write their expansion explicitly:
\begin{equation}
\overline{\Psi}_{2k+1}
=\sum_{n=k+1}^{\infty}B_{2k+1}^{(n)}\frac{1}{y^{2n-1}}
\,, \qquad
\overline{\psi}_{2k}
=\sum_{n=k+1}^{\infty}b_{2k}^{(n)}\frac{1}{y^{2n-1}} \,.
\label{eq:ppy2}
\end{equation}
Substituting these and the expansions (\ref{eq:ppy}) for
$\psi_n^{(0)}$ and $\Psi_n^{(0)}$ into (\ref{e:epscor}) and
(\ref{e:epscor2}), it is possible to solve for the coefficients
$b_k^{(n)}$ and $B_k^{(n)}$, up to one free parameter. Comparing to
(\ref{eq:Psi1ser})-(\ref{eq:psi2ser}) it is natural to choose this
free parameter to be $b_0^{(2)}=c_{-3}$. Similarly to that case,
$b_0^{(2)}$ will only be determined by the requirement that the
extension to the real axis represent a localized
solution. Furthermore, leaving $C$ a free constant and requiring the
absence of logarithmic terms in the expansion of $\overline{\psi}_{k}$
and $\overline{\Psi}_{k}$ yields exactly the value of $C$ given in
(\ref{eq:cc}).

Eq. (\ref{e:epscor}) is consistent with the asymptotics
\begin{equation}
b_2^{(n)}\sim ik_D\,(-1)^n\,\frac{(2n-2)!}{2^{2n-1}}
\left[1+\mathcal{O}\left(\frac{1}{n^3}\right)\right] ,
\end{equation}
where $k_D$ is some constant.  Since the leading order result for
$A_3^{(n)}$ is given by (\ref{eq:a3asympt}), if $k_D\neq 0$, the
coefficients follow the hierarchy $b_2^{(n)}\gg A_3^{(n-1)}$. In order
to be able to extract the value of $k_D$ we have calculated
$b_2^{(n)}$ by solving the mode equations to high orders in $1/y$,
obtaining
\begin{equation}
k_D=1.640\,\frac{D-1}{Q_D}\,. \label{eq:kd}
\end{equation}
The displayed four digits precision for $k_D$ can be relatively easily
obtained by setting $N_F\geq 4$, $N_e\geq 5$ and calculating
$b_2^{(n)}$ to orders $n\geq 25$. We note that there is also a term
proportional to the unknown $b_0^{(2)}=c_{-3}$ in each $b_2^{(n)}$,
giving a $c_{-3}$ dependent $k_D$. Luckily, the influence of this term
to $k_D$ quickly becomes negligible as $N_F$ and $N_e$ grow, making
the concrete value of $c_{-3}$ irrelevant for our purpose.

The Borel summation procedure can be done similarly as in
Eqs. (\ref{e:borelsum})-(\ref{e:borel}). On the imaginary axis
$\psi_2$ is real to every order in $1/y$, however it gets a small
imaginary correction from the summation procedure given by
\begin{equation}
{\rm Im}\,\widehat{\psi}_2(y)=\varepsilon\,
\frac{k_D\pi}{2}\,\exp\left(-2iy\right)\,. \label{eq:impsi2lin}
\end{equation}

\subsection{Extension to the real axis}

Solutions of the Fourier mode equations (\ref{eq:Psin}) and
(\ref{eq:psin}) can be considered to be the sum of two parts. The
first part corresponds to the result of the small amplitude expansion,
the second to an exponentially small correction to it. The small
amplitude expansion is an asymptotic expansion, it gives better and
better approximation until reaching an optimal order, but higher terms
give increasingly divergent results. The smaller $\varepsilon$ is, the
higher the optimal truncation order becomes, and the precision also
improves. The small amplitude expansion procedure gives time-periodic
localized regular functions to all orders, characterizing the core
part of the quasibreather. Their extension to the complex plane is
real on the imaginary axis. Furthermore, the functions obtained by the
$\varepsilon$ expansion are smooth on large scales, missing an
oscillating tail and short wavelength oscillations in the core
region. On the imaginary axis, to a very good approximation, the small
second part of the solution of the mode equations (\ref{eq:Psin}) and
(\ref{eq:psin}) is pure imaginary, and satisfies the homogeneous
linear equations obtained by keeping only the left hand sides of these
equations, because the quasibreather is a small-amplitude one. In the
''inner region'' it is of order $1/y^2$, while on the real axis its
amplitude is of order $\varepsilon^2$, hence to leading order the
quasibreather core background does not contribute. In the previous
subsection we have determined the behaviour of this small correction
close to the poles. Now we extend it to the real axis.

In the ``inner region'', close to the pole, the function ${\rm
  Im}\,\widehat{\psi}_2$ given by (\ref{eq:impsi2lin}) solves the
homogeneous linear differential equations given by the left hand side
of (\ref{eq:psin2}). The extension of this function to the real axis
will provide the small correction to the small amplitude result mentioned
in the previous paragraph. We intend to find the solution
$\widehat{\psi}_2$ of the left hand side of (\ref{eq:psin}), which
reduces to the value given by (\ref{eq:impsi2lin}) close to the upper
pole, where $r=iQ_{D}/\varepsilon+y$, and behaves as
\begin{equation}
{\rm Im}\,\widehat{\psi}_2(y)=-\varepsilon\,
\frac{k_D\pi}{2}\,\exp\left(2iy\right)\,.
\end{equation}
near the lower pole, where $r=-iQ_{D}/\varepsilon+y$. We follow the
procedure detailed in \cite{moredim}.  The resulting function for
large $r$ is
\begin{equation}\label{eq:p2asimpt}
\widehat{\psi}_2=\varepsilon\,
\frac{ik_D\pi}{2}\left(\frac{Q_{D}}{\varepsilon r}\right)^{(D-1)/2}
\exp\left(-\frac{2Q_{D}}{\varepsilon}\right)
\left[i^{(D-1)/2}\exp(-2ir)
-(-i)^{(D-1)/2}\exp(2ir)\right] .
\end{equation}
The general solution of the left hand side of (\ref{eq:psin}) can be
written as a sum involving Bessel functions $J_n$ and $Y_n$, which
have the asymptotic behaviour
\begin{eqnarray}
J_{\nu}(x)&\to&\sqrt{\frac{2}{\pi x}}\,\cos
\left(x-\frac{\nu\pi}{2}-\frac{\pi}{4}\right) ,\\
Y_{\nu}(x)&\to&\sqrt{\frac{2}{\pi x}}\,\sin
\left(x-\frac{\nu\,\pi}{2}-\frac{\pi}{4}\right) ,
\end{eqnarray}
for $x\to+\infty$. The solution satisfying the asymptotics given by
(\ref{eq:p2asimpt}) is
\begin{equation}\label{eq:yasimpt}
\widehat{\psi}_2=\sqrt{\pi}\,
\frac{\alpha_D^{}}{r^{D/2-1}}Y_{D/2-1}(2r) \,,
\end{equation}
where the amplitude at large $r$ is given by
\begin{equation}\label{eq:tampl}
\alpha_D^{}=\varepsilon
\pi k_D\left(\frac{Q_{D}}{\varepsilon}\right)^{(D-1)/2}
\exp\left(-\frac{2Q_{D}}{\varepsilon}\right) .
\end{equation}
For $D>2$ the function given by (\ref{eq:yasimpt}) is singular at the
center, due to the usual central singularity of spherical waves.
Since the amplitude of the quasibreather core is proportional to
$\varepsilon^2$, and its size to $1/\varepsilon$, for small
$\varepsilon$ it is possible to extend the function $\widehat{\psi}_2$
in its form (\ref{eq:yasimpt}) to the real axis into a region which is
outside the domain where $\widehat{\psi}_2$ gets large, but which is
still close to the center when considering the enlarged size of the
quasibreather core. When extending this function further out along the
real $r$ axis, because of the large size of the quasibreather core,
the nonlinear source terms on the right hand side of (\ref{eq:psin})
are not negligible anymore, and the expression (\ref{eq:yasimpt}) for
$\widehat{\psi}_2$ cannot be used. What actually happens is that
$\widehat{\psi}_2$ tends to zero exponentially as $r\to\infty$. This
follows from the special choice of the ``inner solution'' close to the
singularity; namely, we were looking for a solution which agreed with
the small amplitude expansion for ${\rm Re}\,y\to\infty$. The small
amplitude expansion gives exponentially localized functions to each
order and we also required decay beyond all orders when choosing the
contour of integration in the Borel summation procedure.

By the above procedure we have constructed a solution of the mode
equations which is singular at $r=0$. The singularity is the
consequence of the initial assumption of exponential decay for large
$r$. The asymptotic decay induces an oscillation given by
(\ref{eq:yasimpt}) in the intermediate core, and a singularity at the
center. In contrast, the quasibreather solution has a regular center,
but contains a minimal amplitude standing wave tail
asymptotically. Considering the left hand side of (\ref{eq:psin}) as
an equation describing perturbation around the asymptotically decaying
solution, we just have to add a solution $\delta\psi_2$ determined by
the amplitude (\ref{eq:tampl}) with the opposite sign of
(\ref{eq:yasimpt}) to cancel the oscillation and the singularity in
the core. This way one obtains the regular quasibreather solution, whose minimal
amplitude standing wave tail is given as
\begin{eqnarray}\label{eq:qbasimpt}
\phi_{QB}&=&-\sqrt{\pi}\,
\frac{\alpha_D^{}}{r^{D/2-1}}Y_{D/2-1}(2r)\cos(2t) \\
&\approx& -\frac{\alpha_D^{}}{r^{(D-1)/2}}
\sin\left[2r-(D-1)\frac{\pi}{4}\right]\cos(2t) .
\end{eqnarray}
Adding the regular solution, where $Y$ is replaced by
$J$, would necessarily increase the asymptotic amplitude.

If we subtract the incoming radiation from a QB and cut the remaining
tail at large distances, we obtain an oscillon state to a good
approximation.  Subtracting the regular solution with a phase shift in
time, we cancel the incoming radiating component, and obtain the
radiative tail of the oscillon,
\begin{eqnarray}
\phi_{osc}&=&-\sqrt{\pi}\,
\frac{\alpha_D^{}}{r^{D/2-1}}\left[Y_{D/2-1}(2r)\cos(2t)
-J_{D/2-1}(2r)\sin(2t)\right]\\
&\approx& -\frac{\alpha_D^{}}{r^{(D-1)/2}}
\sin\left[2r-(D-1)\frac{\pi}{4}-2t\right] .
\end{eqnarray}
The radiation law of the oscillon is easily obtained now,
\begin{equation}\label{e:radlaw}
\frac{\mathrm{d} E}{\mathrm{d} t}=-k_D^2\pi^2\,
\frac{4\pi^{D/2}}{\Gamma\left(\frac{D}{2}\right)}\,\varepsilon^2	
\left(\frac{Q_{D}}{\varepsilon}\right)^{D-1}
\exp\left(-\frac{4Q_{D}}{\varepsilon}\right)	,
\end{equation}
where the constant $k_D$ is given by (\ref{eq:kd}). If we assume
adiabatic time evolution of the $\varepsilon$ parameter determining
the oscillon state, using Eqs. (\ref{e:energyeps}) and
(\ref{e:energyeps2}) giving $E$ as a function of $\varepsilon$, we get
a closed evolution equation for small amplitude oscillons, determining
their energy as the function of time.

For the physically most interesting case, $D=3$ we write the evolution
equation for $\varepsilon$ and its leading order late time behavior
explicitly:
\begin{eqnarray}
\frac{\mathrm{d} \varepsilon}{\mathrm{d} t}&=&-30.29\,
\exp\left(-\frac{15.909}{\varepsilon}\right)\\
\varepsilon&\approx& \frac{15.909}{\ln t} \,,\quad 
E\approx \frac{1401.6}{\ln t}\,.	
\end{eqnarray}

\section{Acknowledgments}

This research has been supported by OTKA Grants No. K61636,
NI68228.

\end{document}